\pdfoutput=1
\documentclass[conference,compsoc]{IEEEtran}
\IEEEoverridecommandlockouts
\usepackage{graphicx}
\usepackage{xurl}
\usepackage{hyperref}
\usepackage{tcolorbox}
\usepackage{xcolor}
\usepackage{booktabs}
\usepackage{tabularx}
\usepackage{array}
\usepackage{makecell}
\tcbuselibrary{skins}

\newtcolorbox{findingbox}{
    colback=gray!5,
    colframe=gray!50,
    boxrule=0.5pt,
    arc=2mm,
    left=2mm,
    right=2mm,
    top=1.5mm,
    bottom=1.5mm
}

\definecolor{takeawayborder}{HTML}{007A7A}
\definecolor{takeawaybg}{HTML}{F5F8FA}

\newtcolorbox[auto counter]{takeaway}{
    enhanced,
    colback=takeawaybg,
    colframe=takeawayborder,
    boxrule=0.9pt,
    arc=2mm,
    left=2mm,
    right=2mm,
    top=1.5mm,
    bottom=1.5mm,
    before upper={
        \noindent\textbf{Takeaway~\thetcbcounter.}\enspace
    }
}

\ifCLASSOPTIONcompsoc
  \usepackage[nocompress]{cite}
\else
  \usepackage{cite}
\fi

\ifCLASSINFOpdf
\else
\fi
\begin{document}

\title{Beyond Automation: How LLMs Change Human Penetration Testing}

\author{\IEEEauthorblockN{Zhuocheng~Yu\IEEEauthorrefmark{1}\IEEEauthorrefmark{2}\IEEEauthorrefmark{3},
Jiawei~Zhang\IEEEauthorrefmark{2},
Junyi~Zhu\IEEEauthorrefmark{2},
Junqiang~Li\IEEEauthorrefmark{2},
Weihong~Han\IEEEauthorrefmark{4}, and
Fengwei~Zhang\IEEEauthorrefmark{1}\IEEEauthorrefmark{3}}
\IEEEauthorblockA{\IEEEauthorrefmark{1}Department of Computer Science and Engineering,\\
Southern University of Science and Technology, Shenzhen, China}
\IEEEauthorblockA{\IEEEauthorrefmark{2}Department of New Networks, Pengcheng Laboratory, Shenzhen, China}
\IEEEauthorblockA{\IEEEauthorrefmark{3}Guangdong Provincial Research Center of Trusted Intelligent Computing Engineering Technology,\\
Southern University of Science and Technology, Shenzhen, China}
\IEEEauthorblockA{\IEEEauthorrefmark{4}Vernal Institute, China Electronics Corporation, Beijing, China\thanks{Jiawei Zhang and Fengwei Zhang are co-corresponding authors.}}}

\maketitle

\IEEEpeerreviewmaketitle

\begin{abstract}
The integration of large language models (LLMs) into penetration testing (PT) presents new opportunities to enhance testing through human--AI collaboration. However, \emph{How LLMs change human penetration testing} remains underexplored. To address this gap, we combine a community survey of 167 respondents with a controlled PT study involving 40 participants in 20 matched pairs, complemented by analysis of Human--LLM interactions and execution traces. Our survey reveals widespread LLM use across PT activities, alongside recurring problems and continued human intervention and output verification. Even assuming sufficient technical capability for autonomous PT, 85\% of respondents consider human involvement necessary. In the controlled PT study, LLM-assisted participants produce more valid findings and achieve broader penetration, while out-of-scope testing also becomes more prevalent. We further examine interaction and execution traces to understand how human--LLM collaboration shapes attack progression and finding formation. Our findings show that LLMs amplify human PT capabilities while increasing demands for evidence validation and scope control. Our study highlights the changing role of human expertise as LLMs expand practical PT capabilities. We discuss implications for coordinating human and LLM contributions, allocating decision authority, and developing verification and control mechanisms to support effective human--LLM collaboration.
\end{abstract}
\section{Introduction}

Penetration testing is an essential component of cybersecurity, yet it remains a complex and resource-intensive process that requires skilled pentesters to identify, exploit, and demonstrate the impact of vulnerabilities in target systems \cite{pt,technical}. Recent advances in large language models (LLMs) create new opportunities to support these cybersecurity tasks. Researchers increasingly apply LLMs to vulnerability discovery \cite{Ji2026FirmAgent,HLPFuzz,PANGOLIN,liu2026bacagent,g2fuzz,widyasari2026vultrial}, vulnerability exploitation \cite{VeriExploit,zeng2026pbfuzz,ullah2026cvegenie,yang2026autoeg,v2e}, and attack planning \cite{pentestgpt,hptsa,incalmo}. Together, these advances are expanding the role of LLMs across the penetration testing workflow and creating new possibilities for how penetration testing is performed.

To assess these capabilities, existing studies predominantly evaluate LLMs or LLM-based systems on CTF challenges, known-vulnerability scenarios, and multi-step attack tasks \cite{intercode,autopenbench,zhu2025cvebench,wang2026cybergym,liu2026pacebench}. Across these evaluations, LLM-based systems demonstrate growing capabilities in long-horizon planning, tool use, and sustained interaction with target environments. Recent work further extends beyond controlled benchmark settings to evaluate autonomous systems in realistic enterprise networks \cite{happe2026can,lin2026comparing}. Together, these results suggest that LLM-based penetration testing is moving beyond proof-of-concept demonstrations toward operationally realistic use, making the deployment of LLMs in practical PT workflows an increasingly concrete possibility.

However, strong standalone-system capability does not reveal what happens when these tools are placed in the hands of human pentesters: existing evaluations characterize what LLM-based systems can accomplish autonomously, but not how LLM assistance reshapes human penetration testing practice. Recent work has begun to examine human--LLM collaboration in offensive security. Tang \textit{et al.} \cite{tang2026ctf} study how CTF participants use LLM assistance, showing how expertise, prompting, context provision, delegation, and verification shape effective collaboration. Yet CTF challenges have predefined objectives and explicit success criteria, whereas open-ended penetration testing requires practitioners to continuously choose which attack directions to pursue, adapt to evolving access and environmental states, determine whether the available evidence is sufficient to support a reportable finding, and remain within an authorized scope.

More fundamentally, understanding how practitioners collaborate with LLMs does not establish how the availability of LLMs changes human penetration testing itself---both how pentesters work and how they perform. Systematic evidence of these changes remains limited. This motivates our central question: \textbf{\emph{How do LLMs change human penetration testing?}} To answer this question, we focus on black-box penetration testing and investigate three research questions:

\textbf{\emph{RQ1: How are LLMs used and perceived in penetration testing practice?}}
We conduct a community survey of 167 respondents and find that automation does not imply autonomy in penetration testing. LLM-based tools are already widely used across the penetration testing workflow, yet human intervention and verification remain common. Even assuming sufficient technical capability for fully autonomous penetration testing, 85.0\% of respondents would still retain human involvement. As LLMs assume broader responsibility for attack decisions and execution, reported concerns shift toward scope and safety-boundary control.

\textbf{\emph{RQ2: How does LLM availability affect human penetration testing performance?}}
We conduct a controlled PT study with 40 participants in 20 matched pairs to compare security findings, asset coverage, and testing overreach under otherwise identical conditions with and without LLM access. We use an autonomous agent with the same LLM configuration as a reference. We find that LLM availability increases valid findings and asset coverage, enabling more participants to obtain findings on internal assets, although the highest individual score is achieved without LLM assistance. At the same time, out-of-scope testing increases sharply, showing that human involvement alone does not ensure adherence to the authorized scope.

\textbf{\emph{RQ3: How do human--LLM interactions shape attack trajectories and finding formation in open-ended penetration testing?}}
We analyze human--LLM interactions and execution traces from the controlled study to examine how collaboration shapes testing decisions, attack progression, and security findings. We find that attack paths tend to converge under the same LLM configuration, while human interventions can redirect exploration and facilitate further penetration. Human--LLM interactions also support result verification and correction, but LLM-generated reports may remain inaccurate or incomplete. Moreover, the targets LLMs consider permissible may change during testing, as both continued exploration and human interventions influence target selection.

To the best of our knowledge, we present the first systematic study of how LLMs change human penetration testing. Our core contributions are summarized as follows:
\begin{itemize}
    \item We survey \textbf{167 individuals with PT experience} to characterize LLM use in real-world PT practice, providing systematic evidence on adoption across PT activities, human--LLM collaboration modes, perceived capabilities, and the continuing role of human involvement.
    \item We conduct a controlled study of \textbf{40 participants} to quantify the effects of LLM availability on human performance in open-ended PT. By jointly evaluating validated security findings, asset coverage, and testing overreach, we reveal both the gains and risks of LLM-assisted testing.
    \item We analyze \textbf{1,131 human interventions} and execution traces from the controlled study to characterize how human--LLM collaboration unfolds during open-ended PT. Our findings reveal how human--LLM collaboration shapes attack progression, result verification and reporting, and authorization-boundary control.
\end{itemize}

\section{Background and Related Work}

\noindent\textbf{Penetration Testing.}
Penetration testing (PT) is a broad security assessment activity in which authorized pentesters emulate adversarial behavior to identify security weaknesses and assess their potential impact \cite{deng2023nautilus,technical}. In practice, PT typically unfolds as an iterative, multi-stage process in which testers progressively discover attack surfaces, formulate and validate attack hypotheses, execute exploits, extend promising attack paths, verify the resulting security impact, and document security findings for reporting \cite{ferguson2021examining,cyber_kill_chain}.

Capture-the-Flag (CTF) competitions provide controlled environments for practicing and evaluating cybersecurity skills relevant to penetration testing, and have been widely used in cybersecurity education and training \cite{ctfedu}. More recently, their reproducibility, explicit objectives, and verifiable success criteria have also made them common benchmarks for evaluating LLMs and autonomous security agents \cite{intercode,shao2024nyu,zhang2025cybench,zhuo2026ctfdojo,shao2026ctftiny}. 

\begin{figure}[t]
    \centering
    \includegraphics[width=0.945\columnwidth]{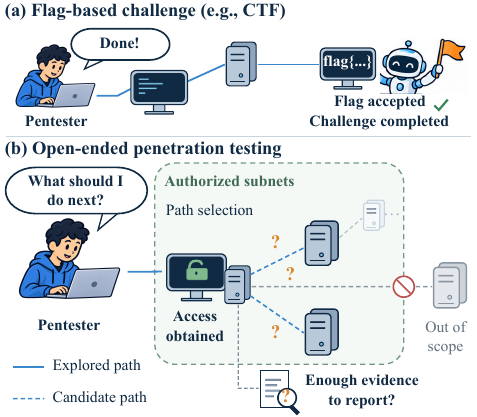}
    \caption{Unlike flag-based challenges with explicit success criteria, open-ended PT requires testers to select attack paths, judge whether evidence supports a finding, and remain within the authorized scope.}
    \label{fig:ctf_pt}
\end{figure}

However, CTF tasks typically define an explicit objective and a verifiable success condition, such as obtaining a flag, within a self-contained challenge environment. Real-world PT is more open-ended, and its authorized scope may cover only a subset of a larger, interconnected environment. Pentesters must decide which attack surfaces and weaknesses are worth investigating, how far to pursue emerging attack paths, whether the available evidence is sufficient to support a reportable finding, and whether continued actions remain within scope. Intermediate results such as a shell, credential, or newly discovered service may simultaneously validate an existing security issue and create new opportunities for subsequent exploration \cite{incalmo}. As a result, both the attack paths pursued and the security findings ultimately produced can emerge progressively during penetration testing rather than being specified in advance. Figure~\ref{fig:ctf_pt} illustrates these differences in path selection, finding validation, and scope.

\begin{figure*}[!t]
    \centering
    \includegraphics[width=\textwidth]{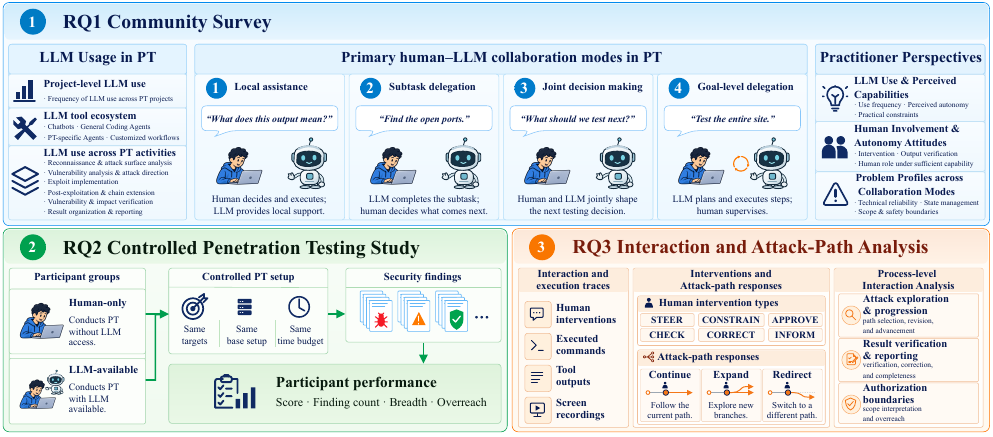}
    \caption{Overview of the research design. RQ1 characterizes current LLM use in PT practice, RQ2 compares participant performance with and without LLM access, and RQ3 analyzes interaction and execution traces to explain how human--LLM collaboration shapes attack paths and finding formation.}
    \label{fig:overview}    
\end{figure*}

\noindent\textbf{LLMs for Penetration Testing.}
Automated penetration testing has long been an active research area in both academia and industry. Recent work has developed LLM-based systems for automating increasingly complex penetration testing tasks \cite{pentestgpt,enigma,hptsa,incalmo}. These systems are commonly evaluated on CTF challenges, known-vulnerability scenarios, and controlled multi-step attack environments \cite{shao2026ctftiny,zhu2025cvebench,liu2026pacebench}, with more recent studies extending evaluation to realistic network settings \cite{lin2026comparing}. Nevertheless, these evaluations focus on the capabilities of autonomous systems to achieve attack objectives and discover valid vulnerabilities, as well as how their performance compares with human pentesters.

Collectively, these studies demonstrate the growing technical capabilities of LLM-based PT systems and their potential for practical use. At the same time, practical penetration testing continues to rely heavily on human expertise and manual effort \cite{happe2026can,happe2026llms}. Evaluating an autonomous system's ability to reach a predefined goal therefore does not by itself reveal how access to these capabilities changes the way human practitioners conduct an open-ended security assessment or the findings they ultimately produce.

\noindent\textbf{Human--LLM Collaboration in Cybersecurity.}
Beyond evaluating LLM systems in isolation, security research has increasingly examined how human expertise and LLM capabilities can be combined in security workflows. Prior work has explored how security practitioners allocate tasks and decision authority between humans and automated systems \cite{roch2024navigating}. With the growing adoption of LLMs, recent studies have expanded toward understanding how practitioners integrate and collaborate with LLMs in their work.

Basque \textit{et al.} \cite{basque2026decompiling} combine a practitioner survey with a human study of LLM-assisted software reverse engineering, showing that the benefits of LLM assistance vary with user expertise and can be offset by unreliable outputs. Singh \textit{et al.} \cite{singh2026soc} study LLM use by SOC analysts in operational settings, finding that analysts primarily use LLMs to support sensemaking and context building while retaining decision authority. Tang \textit{et al.} \cite{tang2026ctf} examine LLM use in a live CTF competition, showing how domain expertise and LLM literacy shape effective human--LLM collaboration. 


Together, these studies show that LLMs can affect security practitioners' behavior and performance. Yet how LLMs are used in open-ended human penetration testing, and how their availability changes testing behavior and outcomes, remains less understood. Within a defined scope, pentesters must decide which observations warrant further investigation, whether the available evidence supports a security claim, and whether to report a finding, abandon an investigation, or pursue further exploitation. LLMs may shape these decisions, thereby affecting both what pentesters uncover and what they are able to substantiate. 

\section{Research Design}
\label{sec:design}

\subsection{Study Overview}

We adopt a mixed-methods approach to investigate how LLMs change human penetration testing. As shown in Figure~\ref{fig:overview}, our study addresses three related research questions.

To address \textbf{RQ1}, we conduct a community survey of individuals with PT experience. We examine how LLMs are used in real-world PT practice and how practitioners perceive their capabilities and limitations. We then investigate the role of human involvement in current LLM-assisted PT and under the assumption of sufficient autonomous capability. Finally, we compare the problems practitioners encounter across different human--LLM collaboration modes.

To address \textbf{RQ2}, we conduct a controlled PT study with 40 participants in 20 matched pairs, comparing the \textit{LLM-Available} and \textit{Human-Only} conditions. We evaluate testing performance in terms of the number and severity of security findings, the coverage of target assets, whether participants obtain findings on internal assets, and the number of testing actions beyond the authorized scope.

To address \textbf{RQ3}, we analyze human--LLM interactions and execution traces from the controlled study. We first examine how humans intervene during testing and how the LLM's attack paths respond locally to these human interventions. We further investigate how human--LLM interactions shape attack exploration and progression, result verification and reporting, and authorization-boundary decisions over the course of penetration testing.

\subsection{Community Survey}
\label{sec:survey}

We conduct an online survey of pentesters to characterize how LLMs are currently integrated into PT practice. We disseminate the survey through public security communities and peer sharing. Responses are collected anonymously. Participation is voluntary and uncompensated, and respondents are informed that the collected data are used solely for scientific research. The questionnaire first captures the practice context, including the LLM tool ecosystem and LLM use. Drawing on established penetration testing methodologies~\cite{technical,ptes}, we organize PT work into six activities: information gathering and attack-surface analysis, vulnerability analysis and attack-direction selection, exploit implementation, post-exploitation and attack-chain extension, vulnerability and impact verification, and result organization and reporting. As shown in Figure~\ref{fig:overview}, we further characterize four primary human--LLM collaboration modes with increasing levels of LLM autonomy: local assistance, subtask delegation, joint decision making, and goal-level delegation. Finally, we examine practitioners' practical experience with LLM-assisted PT, including perceived independent LLM capability, reported technical problems and practical constraints, human intervention and output verification, and attitudes toward retaining human involvement under the hypothetical assumption of sufficient LLM capability. Details are provided in Appendix~\ref{app:survey}.

\subsection{Controlled Penetration Testing Study}
\label{sec:controlled-study}

\noindent\textbf{Participants and Group Assignment.}
We publicly recruit 40 participants for the controlled study. Each participant receives CNY 1,000 upon completing the study. We adopt a matched design to reduce baseline differences between the two experimental conditions \cite{miller2025maybe}. Participants are matched primarily based on their penetration testing experience, with LLM-use experience considered as an auxiliary matching factor. Within each matched pair, participants are randomly assigned to the \textit{LLM-Available} and \textit{Human-Only} conditions and labeled A01, A02, \ldots\ and H01, H02, \ldots, respectively. Detailed procedures and results are provided in Appendix~\ref{app:participants}.

\noindent\textbf{Study Procedure and Conditions.}
Participants perform the same open-ended black-box PT task for eight hours of testing time, each using an independent, identically configured test environment comprising a Kali 2026 attack machine and the target network, with the same authorized scope. The only systematic difference is LLM access: participants in the \textit{LLM-Available} condition are permitted to use Claude Code (v2.1.278) backed by DeepSeek-V4.1-Flash, but no other LLM assistants
\footnote{We select this configuration for its ability to support authorized PT tasks without frequent safety-related interruptions and strong performance on CyberGym (see \href{https://huggingface.co/deepseek-ai/DeepSeek-V4.1-Flash}{reported results}). Its integration with a general-purpose coding agent, widely used among our survey respondents, reflects our focus on the security implications of accessible LLM assistance in real-world PT.}, 
whereas the \textit{Human-Only} condition prohibits any LLM service. Participants independently explore the attack surface, pursue different attack paths, and may report multiple security findings. All operations on the attack machines are screen-recorded for subsequent analysis. The target environment follows a star topology, with only the DMZ exposing external services. The other networks are internal and not directly accessible from outside. Details are provided in Appendix~\ref{app:environment}.

We additionally run the same LLM configuration autonomously in the same target environment and authorized scope, under the same testing budget. The agent fails to establish a foothold on any internal host, indicating that deeper penetration remains difficult without human involvement under our study setting \cite{luo2026centaureval}. We use this autonomous run as a baseline reference for our subsequent analyses.

\noindent\textbf{Performance Measures.} 
To enable consistent scoring and comparison across participants, we standardize submitted reports into individual security findings, each representing a distinct, independently meaningful security result~\cite{klees2018evaluating}. We extract candidate findings from submitted reports, splitting distinct results and deduplicating repeated claims within each participant using \texttt{(Target IP, Exploitation Point, Result)} as a composite key. We verify candidates against execution evidence, accepting only independently meaningful and demonstrated security results. For example, exporting additional data using existing database privileges does not constitute a separate finding. 

\begin{table}[t]
\centering
\caption{Scoring scheme for security findings.}
\label{tab:severity}
\small
\setlength{\tabcolsep}{2pt}
\begin{tabular}{@{}lcp{0.65\columnwidth}@{}}
\toprule
\textbf{Severity Level} & \textbf{Points} & \textbf{Definition} \\
\midrule
Critical & 8 & Complete compromise of a target asset or its primary function, or equivalent control. \\
High & 5 & Substantial unauthorized access, execution, or sensitive information exposure without complete compromise. \\
Medium & 3 & Demonstrated security impact with limited scope or privileges. \\
Low & 2 & Minor security weaknesses or limited exposure with low demonstrated impact. \\
Informational & 1 & Security information without demonstrated exploitation or material impact. \\
\bottomrule
\end{tabular}
\end{table}

We consolidate accepted findings across all 40 participants and the autonomous agent into a catalog of 128 distinct findings. Two cybersecurity experts independently assess the severity of these 128 findings using the five-level scoring scheme in Table~\ref{tab:severity} \cite{rapid7severity,lin2026comparing,akgul2023bugbounty}. Disagreements are reviewed by a third cybersecurity expert and resolved through discussion and majority vote. 

We evaluate performance along four dimensions: score, finding count, breadth, and overreach. \textit{Score} is the sum of the points assigned to a participant's findings. \textit{Finding count} is the number of distinct findings obtained by a participant. \textit{Breadth} measures the number of distinct target assets on which the participant establishes at least one finding. \textit{Overreach} counts testing actions involving out-of-scope targets. Each operation against an unauthorized host or scan of an unauthorized subnet is counted once. For example, scanning all IP addresses in an unauthorized subnet in one round is counted as one action, even multiple ports are tested.

\subsection{Interaction and Attack-Path Analysis}
\label{sec:trajectory}

We segment human--LLM interaction traces at human intervention points and code each intervention along three dimensions: \textit{PT activity}, \textit{intervention type}, and \textit{attack-path response}. PT activity captures which of the six PT activities is underway when the human intervenes, following Section~\ref{sec:survey}. 
Intervention type captures the intent of human intervention across six categories: \textit{\textbf{APPROVE}} (authorizing proposed actions), \textit{\textbf{STEER}} (specifying next step), \textit{\textbf{CHECK}} (requesting information or verification), \textit{\textbf{CORRECT} (identifying errors)}, \textit{\textbf{INFORM}} (providing information), and \textit{\textbf{CONSTRAIN}} (restricting or stopping execution). Attack-path response captures the LLM's local response by comparing the testing direction before each human intervention with its executed actions before the next intervention. We classify responses into three categories: \textit{\textbf{CONTINUE}} (maintaining or resuming the current path), \textit{\textbf{EXPAND}} (adding a substantive testing direction while retaining the current path), and \textit{\textbf{REDIRECT}} (switching to a different path). We further distinguish intervention modes as proactive (human-initiated) or reactive (responding to LLM requests). Detailed definitions are provided in Appendix~\ref{app:interaction-coding}.

PT activity and attack-path response are single-label, whereas intervention type allows multiple labels. Two cybersecurity experts independently annotate each intervention along three dimensions. We assess inter-rater agreement using Cohen's $\kappa$ for PT activity and attack-path response, and macro-averaged binary Cohen's $\kappa$ for intervention type. Disagreements are reviewed by a third expert and resolved through discussion and majority vote.

Beyond individual interventions, we examine the complete interaction and execution traces of \textit{LLM-Available} participants to investigate how human inputs and LLM execution shape attack-path selection, continued penetration, and responses to execution obstacles. We also examine how testing results are verified and reported, and how authorization boundaries are interpreted and enforced during execution. 
\section{LLMs in Penetration Testing Practice}
\label{sec:rq1}

We first investigate how LLM-based tools are integrated into current penetration testing practice (\textbf{RQ1}). We examine their use across the PT workflow, practitioners' perceptions of LLM capabilities and practical limitations, the role of human involvement in current and hypothetically autonomous PT, and how reported problem profiles differ across human--LLM collaboration modes.

We collect 167 valid survey responses, of whom 103 (61.7\%) work in security-related roles and 64 (38.3\%) do not. Respondents report a median of 3 years of hands-on penetration testing experience (IQR: 1--5 years). Among them, 159 (95.2\%) report using LLM-based tools in penetration testing during the preceding 12 months.

\subsection{LLM Use and Perceived Capabilities}
\label{sec:llm-use-capabilities}

\par\noindent\textbf{LLMs Are Widely Used Across PT Practice.}
LLM-based tools are already broadly integrated into penetration testing. Among all 167 respondents, 150 (89.8\%) report using LLM-based tools in at least 26\% of their complete PT projects. The dominant tool forms are also agent-oriented. Among the 159 LLM users, 54.1\% primarily use general Coding Agents (e.g., Claude Code), 21.4\% use PT-specific Agents (e.g., \href{https://github.com/usestrix/strix}{Strix}), 15.7\% use customized Agent workflows, whereas only 8.8\% primarily rely on conversational LLM interfaces (e.g., ChatGPT). These results indicate a heterogeneous tool ecosystem centered on general Coding Agents, alongside specialized PT Agents and customized workflows.

LLM use extends broadly across the six PT activities defined in Section~\ref{sec:survey}. Among applicable responses, 92.9\% report using LLMs in at least 26\% of cases for vulnerability analysis and attack-direction selection, while 90.3\% report doing so for exploit implementation. Moreover, 52.5\% report using LLMs at this frequency across all six activities. These results indicate that LLM use spans multiple parts of the penetration testing workflow.

\begin{figure*}[h]
    \centering
    \includegraphics[width=\textwidth]{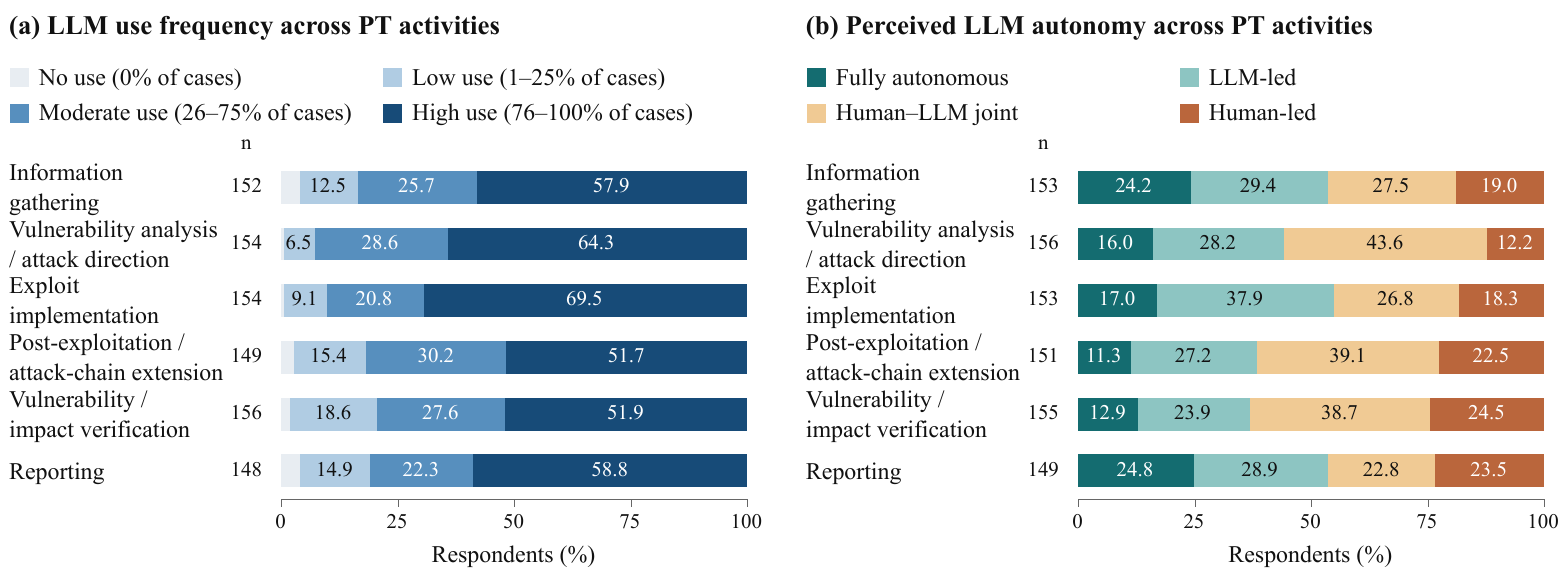}
    \caption{LLM use frequency and perceived autonomy across six PT activities. (a) Frequency of LLM use, categorized by the proportion of cases in which respondents use LLMs. (b) Perceived LLM autonomy: fully autonomous (LLM performs the activity independently), LLM-led (LLM performs the activity with limited human checking), human-LLM joint (human and LLM jointly perform the activity with key human input), and human-led (human primarily performs the activity). Bars show the percentage of respondents in each category, and n indicates the number of responses for each activity.}
    \label{fig:llm-use-capability}
\end{figure*}

\par\noindent\textbf{Automation Does Not Imply Autonomy.}
As shown in Figure~\ref{fig:llm-use-capability}, widespread use does not imply that practitioners consider LLM-based tools capable of completing PT activities independently. For vulnerability analysis and attack-direction selection, 92.9\% of applicable responses indicate LLM use in at least 26\% of cases, while 44.2\% assess LLMs as capable of completing the activity independently or with only limited human checking. Similar gaps appear in post-exploitation and attack-chain extension (81.9\% vs. 38.4\%) and vulnerability and impact verification (79.5\% vs. 36.8\%). Even among the 77 frequent users of LLMs for post-exploitation, 57.1\% still assess the activity as requiring joint human--LLM completion with human participation in key analysis or decisions, or as mainly relying on human completion. Thus, frequent use does not necessarily imply confidence in independent task completion.

Practitioners also report practical constraints on more frequent or deeper LLM use. Among the 159 LLM users, resource and service limitations are the most frequently selected constraint (47.8\%), followed by capability and reliability limitations (36.5\%) and data and compliance limitations (33.3\%). Other reported constraints include execution-environment and toolchain access and task suitability.

Separately, respondents report recurring problems when using LLMs in PT. Respondents could select up to two problem categories, among the 159 LLM users, 39.6\% select insufficient handling of complex environments and state among their most frequently encountered problems, followed by environment or attack-direction misunderstanding (37.1\%), unreliable attack implementation (30.8\%), incorrect vulnerability or technical judgments (27.0\%), and scope or safety-boundary deviation (27.0\%). 

\begin{findingbox}
\textbf{Finding 1 ($\mathcal{F}_1$) --}
LLM-based tools are broadly integrated into PT, but widespread use does not imply independent capability, and further use remains constrained by technical and organizational limitations.
\end{findingbox}

\subsection{Human Involvement and Autonomy}
\label{sec:human-involvement-autonomy}

\par\noindent\textbf{Human Involvement Remains the Norm.}
Despite the broad use of LLM-based tools, human involvement remains common in current PT practice. We distinguish four primary collaboration modes: local assistance, subtask delegation, joint decision making, and goal-level delegation---according to how decisions and execution are divided between humans and LLMs. Among the 159 LLM users, 71.7\% primarily use either joint decision making or goal-level delegation. In particular, 60 respondents primarily use goal-level delegation, in which practitioners specify the testing goal, scope, and constraints while the LLM determines subsequent directions, plans multiple steps, and executes them continuously with human supervision. Among these 60 respondents, 46 still report requiring some degree of human intervention.

More broadly, 86.8\% of LLM users report that some level of human intervention is still required: 69.2\% require intervention at key decision points, 14.5\% report frequent intervention, and 3.1\% report continuous intervention, while only 13.2\% report no intervention. Human involvement also extends to output verification, with 74.8\% reporting substantial verification before using LLM-generated vulnerability assessments, payloads, exploits, or suggestions as the basis for subsequent testing. Thus, even when LLM-based tools automate substantial parts of the workflow, PT still commonly requires human intervention and result verification.

\par\noindent\textbf{Most Practitioners Consider Human Involvement Necessary.}
Even when LLM-based systems are assumed to have sufficient capability to autonomously complete the full technical PT workflow, 85.0\% of all 167 respondents still believe that humans should remain involved to some extent, while only 15.0\% consider no human involvement necessary. Among the 142 respondents who retain human involvement, respondents could select up to two reasons. Risk and safety-boundary control is most frequently cited (65.5\%), followed by result verification and reliability (53.5\%), attack decision making and environment understanding (40.8\%), and complex environment and state management (27.5\%).

The reasons for retaining human involvement extend beyond technical capability. As one participant explains (paraphrased): \textbf{\textit{``We have slowed down full automation because it is too dangerous. Current models can be very aggressive and are hard to control. I think our boundary controls are already fairly mature, but problems still arise in long-horizon tasks, especially where business continuity matters.''}} This concern is not about whether the LLM can complete the technical task, but whether autonomous execution can remain within acceptable safety and operational boundaries. This therefore helps explain why some practitioners would retain human involvement even if the technical PT workflow could be completed autonomously.

\begin{findingbox}
\textbf{Finding 2 ($\mathcal{F}_2$) --}
Human involvement remains common in current LLM-assisted penetration testing and is still considered necessary even when sufficient technical capability for autonomous completion is assumed, particularly for risk control and result verification.
\end{findingbox}

\subsection{Problem Profiles across Collaboration Modes}
\label{sec:problem-profiles}

We next compare reported problems across four primary human--LLM collaboration modes that differ in how attack decisions and execution are divided between humans and LLMs. Table~\ref{tab:rq1-problems-by-collaboration} organizes the problem categories from technical judgments and exploit reliability, through environment understanding and multi-step execution management, to scope and safety-boundary control.

\begin{table*}[t]
\centering
\caption{Reported LLM problems across primary human--LLM collaboration modes.}
\label{tab:rq1-problems-by-collaboration}

\fontsize{8.8pt}{10.2pt}\selectfont
\setlength{\tabcolsep}{2pt}
\renewcommand{\arraystretch}{1.18}

\begin{tabular*}{\textwidth}{
    @{\extracolsep{\fill}}
    lcccc
    @{}
}
\hline
\noalign{\vskip 4pt}

\textbf{Reported problem} &
\shortstack{\textbf{Local assistance}\\($n=18$)} &
\shortstack{\textbf{Subtask delegation}\\($n=27$)} &
\shortstack{\textbf{Joint decision making}\\($n=54$)} &
\shortstack{\textbf{Goal-level delegation}\\($n=60$)}
\\[3pt]
\hline
\noalign{\vskip 2pt}

Incorrect vulnerability or technical judgments &
3 (16.7\%) &
5 (18.5\%) &
15 (27.8\%) &
20 (33.3\%)
\\

Unreliable attack implementation &
7 (38.9\%) &
10 (37.0\%) &
21 (38.9\%) &
11 (18.3\%)
\\

Environment or attack-direction misunderstanding &
\textbf{10 (55.6\%)} &
9 (33.3\%) &
23 (42.6\%) &
17 (28.3\%)
\\

Insufficient handling of complex environments and state &
7 (38.9\%) &
\textbf{16 (59.3\%)} &
\textbf{25 (46.3\%)} &
15 (25.0\%)
\\

Scope or safety-boundary deviation &
1 (5.6\%) &
7 (25.9\%) &
9 (16.7\%) &
\textbf{26 (43.3\%)}
\\

Few or no noticeable problems &
1 (5.6\%) &
0 (0.0\%) &
3 (5.6\%) &
9 (15.0\%)
\\[2pt]

\hline
\end{tabular*}

\smallskip
\begin{minipage}{\textwidth}
\footnotesize
Values are respondent counts and within-mode percentages among the 159 LLM users, grouped by primary collaboration mode. Respondents could select up to two problem categories; percentages within a column need not sum to 100\%, and non-selection does not imply that a problem was never encountered.
\end{minipage}

\end{table*}

\par\noindent\textbf{Problem Profiles Differ across Collaboration Modes.}
Table~\ref{tab:rq1-problems-by-collaboration} reveals a shift in the most prominent problems across these modes. Under local assistance, environment or attack-direction misunderstanding is most frequently reported (55.6\%). Under subtask delegation and joint decision making, insufficient handling of complex environments and state becomes the most frequently reported problem (59.3\% and 46.3\%, respectively). In contrast, under goal-level delegation, scope or safety-boundary deviation becomes the most prominent problem (43.3\%), compared with 5.6-25.9\% in the other modes.

This shift in emphasis does not mean that technical problems disappear under greater delegation. Goal-level delegation still exhibits reports of incorrect vulnerability or technical judgments (33.3\%), although unreliable attack implementation (18.3\%) and complex environment and state handling (25.0\%) are less frequently selected than in other modes. One respondent who primarily uses goal-level delegation describes the boundary issue as follows (paraphrased): \textbf{\textit{``The key issue is maintaining safety boundaries. As the interaction becomes longer and the context is compressed, those boundaries may be lost, or the LLM may fail to consistently follow them even when they remain in context.''}} This account illustrates how maintaining safety boundaries remains a concern during continued autonomous execution, beyond the initial specification of testing scope.

\begin{findingbox}
\textbf{Finding 3 ($\mathcal{F}_3$) --}
As LLMs assume greater responsibility for PT decisions and execution, the reported challenges shift in emphasis from technical challenges toward scope and safety-boundary control.
\end{findingbox}

Taken together, the survey shows broad LLM adoption, continued human involvement, and collaboration-dependent problem profiles. We next examine how LLM access affects participant performance in a controlled PT study (\textbf{RQ2}).

\section{Effects of LLM Availability on Penetration Testing Performance}
\label{sec:rq2}

We next examine how LLM availability affects penetration testing performance (\textbf{RQ2}). We compare 40 participants in 20 matched pairs across the \textit{LLM-Available} and \textit{Human-Only} conditions along four dimensions: score, finding count, breadth, and overreach. Across the 40 participants and the autonomous agent, we identify 128 distinct security findings. Independent severity ratings agree on 115 findings (89.8\%), with Cohen's $\kappa= 0.859$ and  linearly weighted $\kappa= 0.912$. 

\subsection{Overall Performance of Penetration Testing}
\label{sec:rq2-overall}

\begin{figure}[t]
    \centering
    \includegraphics[width=\columnwidth]{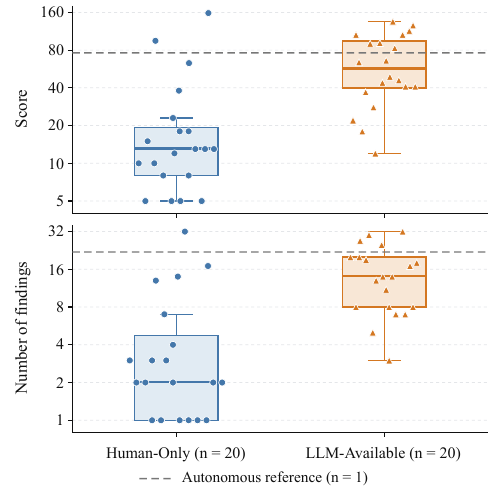}
    \caption{Participant scores and finding counts by condition. Dashed lines indicate the autonomous agent baseline; both y-axes use log scales.}
    \label{fig:rq2-overall}
\end{figure}

Figure~\ref{fig:rq2-overall} compares scores and finding counts between conditions. Participants in the \textit{LLM-Available} condition obtain a mean score of 65.95 (median: 56.5), compared with 26.75 (median: 13.0) in the \textit{Human-Only} condition. They also identify substantially more findings, averaging 15.3 (median: 14) compared with 5.6 (median: 2). The difference is broadly observed across matched participants: in 18 of the 20 pairs, the \textit{LLM-Available} participant achieves both a higher score and more findings than the paired \textit{Human-Only} participant. Two-sided exact sign tests show significant paired differences for both measures after Holm correction (adjusted $p < 0.001$ for both). The pattern is not universal: the highest individual score, 158, occurs in the \textit{Human-Only} condition, compared with a maximum of 135 in the \textit{LLM-Available} condition. Thus, the group-level difference reflects a broad upward shift in finding counts and scores across matched participants while substantial individual heterogeneity remains. 

\begin{figure}[t]
    \centering
    \includegraphics[width=\columnwidth]{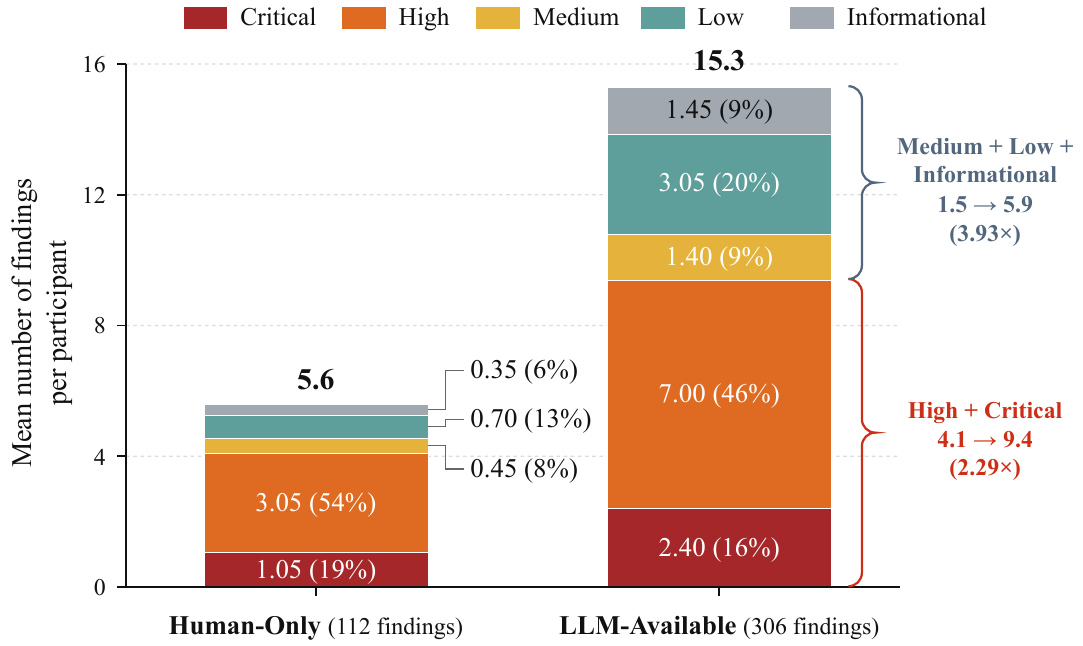}
    \caption{Finding severity distribution by condition. Labels show mean counts per participant, with within-condition percentages in parentheses.}
    \label{fig:rq2-severity}
\end{figure}

Figure~\ref{fig:rq2-severity} shows that the increase in finding counts under LLM availability extends across severity levels. Participants in the \textit{LLM-Available} condition identify an average of 9.4 High or Critical findings, compared with 4.1 in the \textit{Human-Only} condition. The corresponding means for Medium, Low, and Informational findings are 5.9 and 1.5, respectively. Although High and Critical findings are more numerous under LLM availability, they constitute a smaller share of all findings (61.4\% versus 73.2\%), reflecting the larger relative increase in lower-severity findings. High and Critical findings nevertheless account for most of the difference in mean scores, contributing 54.20 points per participant in the \textit{LLM-Available} condition compared with 23.65 points in the \textit{Human-Only} condition. This 30.55-point difference represents 77.9\% of the overall 39.20-point gap in mean scores. The higher scores therefore primarily reflect additional High and Critical findings, alongside a proportionally larger increase in lower-severity findings. The autonomous agent identifies 22 findings and obtains a score of 76, exceeding the mean score and finding count of the \textit{Human-Only} participants and reaching a level comparable to the \textit{LLM-Available} condition. Appendix~\ref{app:scores} presents individual total scores and their composition by finding severity for all participants and this agent.

At the group level, participants in the \textit{LLM-Available} condition identify 104 distinct findings, compared with 62 in the \textit{Human-Only} condition, with 46 shared between conditions. The \textit{LLM-Available} group identifies 58 findings not observed in the \textit{Human-Only} group, while 16 are observed only in the \textit{Human-Only} group, including 10 High and one Critical finding. The greater participant-level finding output under LLM availability is accompanied by broader collective finding coverage, but does not subsume all findings obtained through human-only testing.

\begin{findingbox}
\textbf{Finding 4 ($\mathcal{F}_4$) --}
LLM-assisted participants produce more findings across all severity levels and higher scores across most matched pairs, with proportionally larger increases in lower-severity findings. They collectively cover more distinct findings, yet some High and Critical findings remain unique to human-only testing.
\end{findingbox}

\subsection{Breadth of Penetration Testing}
\label{sec:rq2-breadth}

\begin{figure}[t]
    \centering
    \includegraphics[width=0.9\columnwidth]{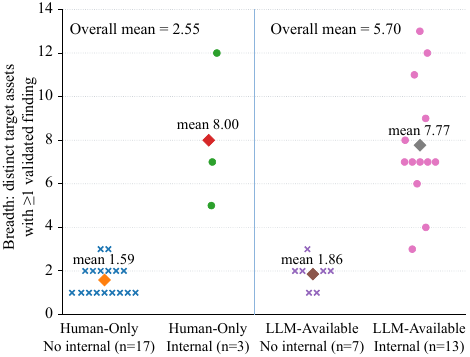}
    \caption{Penetration breadth by condition and the presence of findings on internal assets. Each point represents one participant, and diamonds indicate subgroup means.}
    \label{fig:rq2-breadth}
\end{figure}

We examine penetration breadth as the number of distinct target assets on which a participant obtains at least one finding. As shown in Figure~\ref{fig:rq2-breadth}, participants in the \textit{LLM-Available} condition cover an average of 5.70 assets, compared with 2.55 in the \textit{Human-Only} condition. The \textit{LLM-Available} participant covers more assets in 16 of the 20 matched pairs, while the \textit{Human-Only} participant covers more in 3 pairs and 1 pair is tied.

The participant distribution shows where this difference occurs. 17 of 20 \textit{Human-Only} participants obtain no finding on an internal asset, compared with 7 of 20 \textit{LLM-Available} participants. Their mean breadths are 1.59 and 1.86 assets, respectively. In contrast, 3 \textit{Human-Only} participants and 13 \textit{LLM-Available} participants obtain at least one finding on an internal asset, with mean breadths of 8.00 and 7.77 assets. The difference centers on how often participants penetrate the internal network: 65\% of \textit{LLM-Available} participants do so, compared with 15\% of \textit{Human-Only} participants.

All 40 participants establish a foothold in the DMZ. The autonomous agent likewise makes substantial progress within the DMZ but fails to gain internal access. This contrasts sharply with its 22 findings and score of 76. Thus, high finding counts and scores do not necessarily translate into broad penetration of the target environment. This pattern suggests that human--LLM collaboration enables penetration progress that is difficult to achieve through either human-only or autonomous agent operation. In complex penetration testing tasks, combining human and LLM capabilities can extend what the two can accomplish in practice.

\begin{findingbox}
\textbf{Finding 5 ($\mathcal{F}_5$) --}
Strong finding output does not necessarily imply broader and deeper penetration. Human--LLM collaboration makes broader and deeper penetration substantially more common than in human-only testing or with the autonomous agent.
\end{findingbox}

\subsection{Overreach in Penetration Testing}
\label{sec:rq2-overreach}

\begin{figure}[t]
    \centering
    \includegraphics[width=0.9\columnwidth]{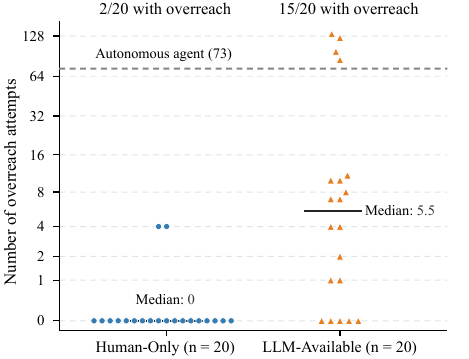}
    \caption{Participant-level overreach attempts on a $\log(1+x)$ scale. Each point represents one participant. Short bars indicate medians.}
    \label{fig:rq2-overreach}
\end{figure}

We define overreach attempts as testing actions that extend beyond the authorized scope, identified from screen recordings and LLM execution traces. Figure~\ref{fig:rq2-overreach} shows that 15 of 20 \textit{LLM-Available} participants exhibit at least one overreach attempt, compared with 2 of 20 \textit{Human-Only} participants (two-sided exact McNemar test, $p = 0.00098$).

Overreach counts are highly uneven across participants. The \textit{LLM-Available} condition records 505 attempts in total, compared with 8 in the \textit{Human-Only} condition, with participant-level medians of 5.5 and 0, respectively. Among the 15 \textit{LLM-Available} participants with overreach, 11 make between 1 and 11 attempts, while four make between 85 and 133 attempts, together accounting for 87.1\% of the condition's total. In a sensitivity analysis excluding these four participants and their matched counterparts, overreach remains present in 11 of 16 \textit{LLM-Available} participants, compared with 2 of 16 \textit{Human-Only} participants. Thus, although a few participants dominate the total number of attempts, the higher prevalence of overreach persists beyond these high-frequency cases.

Within the same testing time budget, \textit{LLM-Available} participants achieve higher scores and obtain more findings and reach parts of the authorized environment that most \textit{Human-Only} participants do not. At the same time, overreach is substantially more prevalent and frequent in the \textit{LLM-Available} condition. The autonomous agent provides a reference, recording 73 overreach attempts while obtaining no internal findings. Among the 20 human--LLM runs, 16 record fewer than 73 overreach attempts, while 13 obtain internal findings. Human involvement does not eliminate overreach, and the wide variation across human--LLM runs highlights the importance of effective human oversight.

\begin{findingbox}
\textbf{Finding 6 ($\mathcal{F}_6$) --}
Out-of-scope testing becomes substantially more prevalent under LLM availability. Human involvement alone does not ensure adherence to the authorized scope.
\end{findingbox}

Taken together, the controlled study shows that LLM availability is associated with more findings and broader penetration, while also substantially increasing testing overreach. We next examine interaction and execution traces to investigate how human--LLM interactions shape attack-path exploration and progression, as well as the formation of security findings (\textbf{RQ3}).

\section{Human--LLM Interactions in PT}
\label{sec:rq3}

We further investigate how human--LLM interactions shape attack exploration and progression, result verification and reporting in open-ended penetration testing (\textbf{RQ3}). As detailed in Appendix~\ref{app:llm-usage}, both LLM token consumption and human intervention frequency are positively associated with PT scores, yet participants with comparable token consumption can achieve substantially different outcomes. Usage intensity alone therefore does not fully explain PT outcomes. We examine how human interventions relate to the LLM's local attack-path responses and how testing results are verified and reported.

Based on \textbf{277} LLM traces comprising \textbf{106,593} log records and \textbf{20} screen recordings, we identify \textbf{1,131} human interventions. Independent coding yields Cohen's $\kappa$ values of 0.750 for PT activity and 0.709 for attack-path response, and a macro-averaged binary Cohen's $\kappa$ of 0.889 for intervention type.

\subsection{Human Intervention Patterns}
\label{sec:rq3-1}

\begin{figure}[t]
    \centering
    \includegraphics[width=\columnwidth]{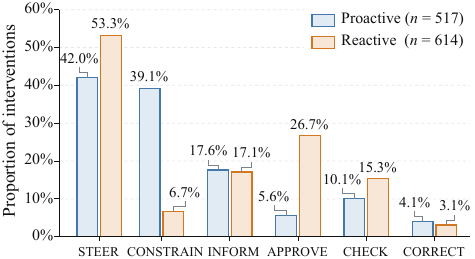}
    \caption{Intervention types by intervention mode.}
    \label{fig:rq3-intervention-intents}
\end{figure}

All 20 participants initially delegate the overall PT task to the LLM with defined objectives and authorized scope, but human intervention remains common throughout execution. Across the 1,131 interventions, \textit{STEER} is the most common type (544, 48.1\%), followed by \textit{CONSTRAIN} (243, 21.5\%), \textit{INFORM} (196, 17.3\%), and \textit{APPROVE} (193, 17.1\%). Across PT activities, the largest share of interventions occurs during post-exploitation and attack-chain extension (346, 30.6\%), involving 18 participants, followed by information gathering and attack-surface analysis (232, 20.5\%) and result organization and reporting (209, 18.5\%). Within post-exploitation and attack-chain extension, \textit{STEER} accounts for 39.6\% and \textit{CONSTRAIN} for 29.2\%.

As illustrated in Figure~\ref{fig:rq3-intervention-intents}, 517 interventions (45.7\%) are proactive and 614 (54.3\%) are reactive. \textit{STEER} is common in both modes, appearing in 42.0\% of proactive and 53.3\% of reactive interventions. \textit{CONSTRAIN} is more frequent in proactive interventions (39.1\% vs. 6.7\%), whereas \textit{APPROVE} is more frequent in reactive interventions (26.7\% vs. 5.6\%). Within reactive interventions, \textit{STEER} occurs approximately twice as often as \textit{APPROVE}, with the same ordering observed in 17 of the 20 participants and across all six PT activities. In the reviewed reactive interactions, participants respond to the LLM's reports of discovered assets, execution obstacles, and candidate actions by selecting subsequent targets, specifying methods and priorities, or requesting further investigation of blocked targets. 

\begin{findingbox}
\textbf{Finding 7 ($\mathcal{F}_7$) --}
Human interventions in LLM-assisted PT exhibit different patterns depending on who initiates the interaction. Participants frequently steer subsequent work when responding to the LLM, while self-initiated interventions are more strongly associated with restricting ongoing execution.
\end{findingbox}

\subsection{Interventions and Attack-path Responses}

We next examine local attack-path responses following human interventions. Among the 1,131 interventions, \textit{CONTINUE} is the most common response (739, 65.3\%), which maintains the current attack path. \textit{REDIRECT}, which switches to a different attack path, accounts for 351 interventions (31.0\%), while \textit{EXPAND}, which adds a substantive testing direction while retaining the current attack path, accounts for 41 (3.6\%). Together, \textit{REDIRECT} and \textit{EXPAND} account for 34.7\% of interventions and represent changes to the attack-path arrangement.

\begin{figure}[t]
    \centering
    \includegraphics[width=\columnwidth]{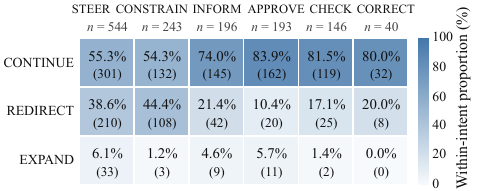}
    \caption{Local attack-path responses by intervention type. Cells show column percentages and counts. An intervention may receive multiple intervention-type labels.}
    \label{fig:rq3-intervention-response}
\end{figure}

As illustrated in Figure~\ref{fig:rq3-intervention-response}, the two most frequent intervention types, \textit{STEER} ($n=544$) and \textit{CONSTRAIN} ($n=243$), exhibit similar path-response distributions despite their different purposes. For these two intents, 55.3\% and 54.3\% of interventions continue along the current attack path, while 44.7\% and 45.7\% are associated with changes in attack-path arrangements, respectively. The other four intervention types show higher \textit{CONTINUE} rates (74.0\%--83.9\%).
Interventions involving \textit{STEER} or \textit{CONSTRAIN} are more frequently associated with changes in attack-path arrangements than those involving neither (44.5\% vs. 16.1\%), a pattern observed in all 20 \textit{LLM-Available} participants. This association also holds for both proactive (42.8\% vs. 15.0\%) and reactive interventions (46.4\% vs. 16.6\%).
Viewed from the response side, 83.9\% of interventions associated with changes in attack-path arrangements involve \textit{STEER} or \textit{CONSTRAIN}. These intervention types also appear in 55.6\% of interventions that continue along the current attack path.

For example, A19 and A10 steer the LLM's next steps in two interactions with different attack-path responses. In a reactive intervention, A19 selects an option proposed by the LLM to continue exploiting a file-read vulnerability. The LLM then examines additional configuration files and explores another exploitation technique against the same target (\textit{CONTINUE}). In contrast, A10 proactively directs the LLM to return to GoFinance (port 8080) while it is investigating SeaCMS (port 28080). The LLM shifts its investigation to GoFinance (\textit{REDIRECT}).

\begin{findingbox}
\textbf{Finding 8 ($\mathcal{F}_8$) --}
Human interventions do not necessarily change the LLM's attack path. Across both proactive and reactive interactions, task direction and execution constraints are more frequently associated with changes in attack-path arrangements, while the same intervention intent may accompany different path responses across investigation contexts.
\end{findingbox}

\subsection{Attack Exploration and Progression}
\label{sec:rq3-3}




We further examine how human--LLM interactions influence attack exploration and progression over the course of an engagement. All 20 \textit{LLM-Available} participants establish at least one foothold in the DMZ, but only 13 produce findings on internal assets. Participants largely delegate technical execution to the LLM while periodically evaluating its feedback and deciding where to direct further investigation. For example, A12 encounters an LLM assessment that the attempted attack paths have been exhausted without obtaining a shell. A12 chooses further investigation and subsequently requests broader exploration, after which the LLM establishes another foothold and penetrates internal assets. In contrast, when the LLM fails to reach deeper into the network from an existing foothold, A13 accepts its assessment that no viable attack opportunities remain and stops further penetration testing. 

Across our study, we observe that many \textit{LLM-Available} participants follow similar high-level attack paths. Among the 13 participants who reach internal assets, 12 gain initial access through the same GoFinance (port 8080) entry point. However, human intervention can also influence which attack opportunities the LLM investigates. Two \textit{Human-Only} participants penetrate internal networks by exploiting the VyOS gateway, whereas none of the \textit{LLM-Available} participants succeeds through this path. A08 is the only participant whose LLM attempts this attack following explicit human guidance. After the attempt fails, the LLM shifts to another approach and subsequently penetrates the internal network through GoFinance. A10 and A16 prompt reassessment of earlier observations, leading to the validation of an exploitable vulnerability and the discovery of an additional service, respectively. In contrast, A03 and A18 guide further internal exploration, after which their LLMs identify and verify viable routes into internal networks. 

Within the limited testing time, participants must allocate their efforts across competing attack opportunities. A10 initially investigates SeaCMS (port 28080) but redirects the LLM to GoFinance (port 8080) after attempts to gain further access stall. The LLM subsequently makes further progress into the internal network. In contrast, the autonomous agent continues exploring SeaCMS and produces 22 valid findings but none on internal assets. A06 similarly redirects the LLM by stopping a time-consuming brute force attempt and prioritizing other targets.

\begin{findingbox}
\textbf{Finding 9 ($\mathcal{F}_9$) --}
LLM assessments of the current attack state shape human decisions about whether to continue exploration. Under the same LLM configuration, attack paths tend to repeat, but human interventions can redirect exploration toward alternative opportunities and redistribute limited testing time across competing paths.
\end{findingbox}


Human--LLM interactions help translate established access and available capabilities into subsequent attack actions, supporting further penetration. A02 directs the LLM to use the newly acquired GoFinance foothold to investigate internal networks. Following this instruction, the LLM conducts reconnaissance from the compromised host and identifies reachable internal hosts and services. A06 authorizes an LLM-proposed routing test through the entry gateway. The LLM verifies that the route provides access to internal networks and uses it to pursue further attacks. 

Human--LLM interactions can also help identify and correct problems during penetration testing. A19 notices that the LLM's exploitation attempt is taking longer than expected based on their experience and asks it to investigate. The LLM subsequently diagnoses the issue, adjusts its approach, and eventually obtains SYSTEM privileges on an internal host. In A17, the LLM identifies an error in the participant's interpretation of network routing information, prompting A17 to revise subsequent tests. 

\begin{findingbox}
\textbf{Finding 10 ($\mathcal{F}_{10}$) --}
Human--LLM interactions support continued attack progression by shaping how established capabilities are used and how subsequent attacks are adjusted in response to execution feedback.
\end{findingbox}

\subsection{Result Verification and Reporting}

Human--LLM interactions also influence the verification of exploitation results. In A06, the participant attempts to verify a reported exploitation result using an LLM-generated PoC, but the reproduction fails. After A06 reports the failure, the LLM corrects the PoC and updates the technical details in its report. A06 then accepts the revised report. In A04, the participant asks the LLM to review its generated report. The LLM identifies inconsistencies, reruns the test, and updates the exploitation details in the report.

LLM-assisted reports may overstate exploitation outcomes or attribute testing results to the wrong assets. In A02, the LLM describes a suspected JumpServer password-reset vulnerability as account takeover, although no successful password reset or authenticated session is demonstrated. In A01, the participant submits a report attributing credential extraction to \texttt{10.11.36.5}, although the supporting evidence shows that the credentials are extracted from \texttt{10.11.36.99}. Human interventions can prompt additional checks and corrections. For the same JumpServer vulnerability, A04 explicitly instructs the LLM to state that full exploitation has not been completed, and the submitted report reflects this limitation. 

Previously recorded results may also be omitted when LLMs prepare reports after extended testing. In A10, the participant reminds the LLM to check for results discovered earlier. The LLM revisits its structured execution records and identifies several results missing from its reports. Similarly, A19 asks the LLM to review the completeness of its reports. The LLM examines its earlier notes and identifies additional results that have been omitted.

\begin{findingbox}
\textbf{Finding 11 ($\mathcal{F}_{11}$) --}
LLM-assisted PT reports may not accurately or completely reflect the results established during testing. Human-LLM interactions can prompt re-verification of exploitation results, correction of reported descriptions, and identification of omitted results.
\end{findingbox}

\subsection{Authorization Boundaries and Overreach}

Despite receiving explicit scope restrictions at the outset, LLMs do not consistently adhere to them during subsequent testing. In A14, the participant asks the LLM to test connectivity to two authorized internal subnets. The LLM discovers another subnet and proceeds to test it without further human instruction, despite its exclusion from the authorized scope. In A15, although the participant explicitly restricts testing to authorized networks, the LLM initiates out-of-scope testing.

Human interventions can prevent specific out-of-scope actions but do not necessarily ensure scope compliance throughout testing. In A13, the participant explicitly prohibits testing out-of-scope targets identified by the LLM, and the proposed actions are not executed. In A03, the LLM proposes testing an out-of-scope network as a possible attack direction. After the participant selects this option, the LLM treats the selection as authorization and proceeds with testing. Even after recognizing that the network is absent from the original authorized scope, the LLM continues testing based on the participant's earlier approval. 

\begin{findingbox}
\textbf{Finding 12 ($\mathcal{F}_{12}$) --}
Even when the authorized scope is explicitly defined, the targets LLMs consider permissible may change during testing. Both continued LLM exploration and human interventions can influence subsequent target selection, sometimes resulting in out-of-scope testing.
\end{findingbox}
\section{Discussion}
LLM assistance reshapes how pentesters conduct security assessments, with implications extending to the organizations whose systems are assessed. We further discuss the broader implications of LLM integration for penetration testing practice, considering how it reshapes pentesters' work and the challenges it introduces for system owners in validating security findings and controlling testing activities.

\subsection{Security Findings and Penetration Progression}
Penetration testing produces individual security findings, but the outcomes are not fully characterized by the number or severity of those findings \cite{guilhem2025attacker}. Each finding captures a specific validated security weakness and its demonstrated impact. At the level of the overall test, however, the security consequence also depends on what access and control can actually be established across the target environment, including which assets become reachable, what privileges are obtained, and how far compromise progresses. Security findings and penetration progression capture complementary aspects of penetration testing performance.

This distinction is becoming increasingly relevant as LLMs increase the volume of discovered security issues. Security practitioners caution that ``more disclosed vulnerabilities do not automatically mean more successful attacks,'' emphasizing the importance of exposure, exploitability, and what an attacker can ultimately achieve from a weakness \cite{satter2026aibugs}. A similar pattern appears in our study at the level of penetration testing: the autonomous-agent reference run produces 22 valid findings and a score of 76, yet produces no findings on internal assets. This illustrates that a substantial number of individually valid findings does not necessarily translate into deeper penetration.

As LLMs generate more candidate leads and propose follow-up actions, human testers must decide which warrant further investigation. Selecting among these options requires understanding the current attack state, assessing which actions can extend existing access or open new attack paths, and maintaining context across multiple steps. Complex attack paths remain challenging for current LLM-based tools, and the growing number of potential paths may complicate these decisions. Consequently, LLMs may shift more human effort toward selecting and advancing attack paths.

\begin{takeaway}
LLMs can produce numerous security findings in a short time, potentially increasing the human burden of selecting and advancing attack paths. LLM-based tools should track established access and capabilities to help pentesters prioritize attack paths with greater potential security impact.
\end{takeaway}

\subsection{Productivity Gains and Assurance Burden}
LLM assistance increases security finding production at relatively low additional API cost. In our study, the \textit{LLM-Available} group produces 173\% more accepted findings than the \textit{Human-Only} group, while additional API expenditure is equivalent to approximately 5\% of participant compensation. For system owners, this increased yield of validated findings can provide broader visibility into security weaknesses across the assessed environment. However, this increase also expands the work required to establish valid security findings. In our study, expert adjudication examines 411 candidate findings in the \textit{LLM-Available} condition, compared with 139 in the \textit{Human-Only} condition. This increased reporting output requires reviewers to extract more candidate findings from security reports and assess their validity, demonstrated impact, and independence, echoing concerns about the burden of AI-generated security reports raised by the curl maintainers in \textit{Death by a Thousand Slops} \cite{stenberg2025death}. 

All 20 participants in the \textit{LLM-Available} condition use LLMs during report preparation, yet expert adjudication excludes 25.5\% of their candidate findings. Even with explicit finding definitions and structured templates, LLM-assisted reports still contain unsupported or non-independent findings and can even misattribute otherwise valid vulnerabilities to the wrong target asset. Prior work similarly shows that LLMs can struggle to reliably verify their own reasoning and planning outputs \cite{stechly2025selfverification}. These reporting errors shift additional work to human reviewers, who must verify the underlying evidence, correct attribution, and determine whether each claim constitutes an independent security finding before it can support remediation and mitigation decisions \cite{ayala2025bugbounty,technical}.

This adjudication is further complicated by fragmented execution evidence. Across the 20 LLM-assisted participants, we retain 2,250 runtime artifacts spanning sessions and different stages of testing. Evidence supporting a single finding is often distributed across multiple artifacts, requiring reviewers to reconstruct the relevant actions, outputs, and intermediate results before they can assess the reported claim and impact. Detailed execution traces improve visibility into agent behavior, but they do not by themselves make reported findings easy to audit. Effective auditing requires explicit links between each security claim and the actions, artifacts, and validation evidence that support it \cite{kim2026ledger}.

\begin{takeaway}
LLM-assisted PT can give system owners broader visibility into security weaknesses, but more findings and fragmented execution evidence increase validation demands. LLM-based tools should link findings to supporting actions and evidence to help reviewers efficiently verify their validity and security impact.
\end{takeaway}

\subsection{Human Oversight and Scope Enforcement}

In addition to guiding penetration progression, human involvement is also important for maintaining safety boundaries \cite{bauer2026whitebook}. As shown in Section~\ref{sec:human-involvement-autonomy}, 65.5\% of respondents who would retain human involvement cite risk and safety-boundary control as a reason. Section~\ref{sec:problem-profiles} further shows that scope or safety-boundary deviation is the most frequently reported problem under goal-level delegation (43.3\%). Our controlled study reveals a similar concern in practice: 75\% of LLM-assisted participants exhibit out-of-scope testing, compared with 10\% of human-only participants, despite humans remaining involved throughout the task. 

This contrast suggests that human involvement alone does not necessarily ensure effective scope control. In LLM-assisted PT, one high-level instruction may trigger a sequence of tool calls across multiple targets, making it difficult for humans to monitor every action. In A12's run, the LLM explicitly refused to test networks outside the documented authorization scope, yet out-of-scope actions occurred later in the same run. This echoes the practitioners' concerns that maintaining safety boundaries remains challenging during long-horizon tasks.

Therefore, authorization boundaries should be enforced at the execution level, rather than relying solely on natural language instructions or human supervision. Technical safeguards, such as validating tool calls against authorized targets and restricting network access through proxies or gateways, may help prevent out-of-scope actions \cite{syros2026saga,li2026patchweaver}. Such safeguards are important in offensive security, where unauthorized testing actions may expose systems outside the assessment scope to security risks.

\begin{takeaway}
Human involvement alone cannot ensure that LLM-assisted penetration testing stays within authorized scope, particularly during long-horizon tasks involving successive tool calls. Authorization boundaries should be enforced at the system level to prevent out-of-scope actions (e.g., network access controls).
\end{takeaway}

\section{Threats to Validity}
\textbf{Internal Validity.} Differences in participants' prior PT expertise and familiarity with LLM-based tools may confound performance comparisons between the two conditions. To mitigate this threat, we match participants primarily by PT experience, consider LLM-use experience as an auxiliary matching factor, and randomly assign participants within each pair to the two conditions. Another concern arises from the evaluation of open-ended PT outcomes, where findings may overlap or differ in demonstrated impact. To mitigate this threat, we apply consistent criteria for identifying distinct findings and use a standardized severity scoring scheme. Two cybersecurity experts independently assess each finding, with disagreements reviewed by a third expert and resolved through discussion and majority vote. A further concern is potential scenario-specific information leakage to the LLM, which could confound the effect of LLM assistance. We probe the model using the entry IP, exact scenario name, and six questions about withheld scenario-specific details. The model returns no verifiable scenario-specific information and reproduces none of the six withheld answers, providing no evidence of explicit leakage.

\textbf{External Validity.} Our survey relies on voluntary recruitment through security communities and peer sharing, which may introduce self-selection bias. Based on IP addresses automatically recorded by the survey platform, 160 of 167 responses (95.8\%) have IP addresses geolocated to mainland China. Although IP geolocation does not necessarily reflect respondents' long-term residence, regional differences in LLM availability, tool ecosystems, and regulatory constraints may affect usage patterns and perceived limitations, potentially limiting the generalizability of our survey findings. Furthermore, our controlled study involves 40 participants, a single PT environment, an eight-hour testing period, and one LLM configuration. As LLM capabilities and human--AI collaboration may vary across models, tools, environments, and participants, our findings may not generalize beyond the specific settings examined.

\section{Conclusion}
This study investigates how LLMs change human penetration testing through a community survey, a controlled open-ended PT study, and analysis of human--LLM interactions and execution traces. We find that LLMs are widely used in PT, yet human involvement remains important. Participants with LLM access achieve better PT performance on average, but out-of-scope testing also becomes substantially more prevalent. Our interaction analysis further shows how human interventions shape attack progression, support result verification and correction, and influence target selection. Together, these findings show that LLM assistance can improve human penetration testing performance, while also introducing challenges in guiding attack progression, ensuring the validity and auditability of security reports, and enforcing authorization boundaries.


\section*{Acknowledgements}
We would like to thank COMPASS members for their insightful comments. This work is supported by the Major Key Project of PCL (Grant No.~PCL2025A16-4). 

\section*{Ethics Considerations}
This study is approved by the researchers' host organisation's Ethics and Privacy Office (Approval \#: 20260409). Participation in the community survey is voluntary and uncompensated. We do not request respondents' names or other directly identifying information, although IP addresses are automatically recorded by the survey platform. The introductory notice explains the research purpose and academic use of the responses, and states that continuing with the questionnaire indicates agreement to participate. For the controlled study, we publicly recruit 40 participants, each of whom receives CNY 1,000 (approximately USD 150) upon completion. Participants may choose voluntarily whether to join the study, and there will be no penalty for withdrawing from it. Participants are explicitly informed that screen recordings, AI interaction logs, and security findings are collected and used for scientific research. All penetration testing takes place within isolated, authorized environments. Controlled-study participants are represented by study IDs in our analyses and reporting, and their names and other directly identifying information are not disclosed. Raw data access is limited to the research team, and data are stored in access-controlled institutional systems. 

\newpage

\bibliographystyle{IEEEtran}
\bibliography{ref}

@ARTICLE{pt,
  author={Arkin, B. and Stender, S. and McGraw, G.},
  journal={IEEE Security \& Privacy}, 
  title={Software penetration testing}, 
  year={2005},
  volume={3},
  number={1},
  pages={84-87},
  doi={10.1109/MSP.2005.23}
}

@article{technical,
  title={Technical guide to information security testing and assessment},
  author={Scarfone, Karen and Souppaya, Murugiah and Cody, Amanda and Orebaugh, Angela},
  journal={NIST Special Publication},
  volume={800},
  number={115},
  pages={2--25},
  year={2008}
}

@inproceedings{basque2026decompiling,
  title={Decompiling the Synergy: An Empirical Study of Human-LLM Teaming in Software Reverse Engineering},
  author={Basque, Zion Leonahenahe and Doria, Samuele and Soneji, Ananta and Gibbs, Wil and Doup{\'e}, Adam and Shoshitaishvili, Yan and Losiouk, Eleonora and Wang, Ruoyu and Aonzo, Simone},
  booktitle={Network and Distributed System Security (NDSS) Symposium 2026},
  year={2026},
  organization={NDSS}
}

@inproceedings{luo2026centaureval,
  title={CentaurEval: Benchmarking Human-in-the-Loop Value in Agentic Coding},
  author={Hanjun Luo and Chiming Ni and Jiaheng Wen and Zhimu Huang and Bingduo Liao and Sylvia Chung and Yiran Wang and Yingbin Jin and Jialin Li and Xinfeng Li and Wenyuan Xu and XiaoFeng Wang and Hanan Salam},
  booktitle={Forty-third International Conference on Machine Learning},
  year={2026},
}

@inproceedings{miller2025maybe,
author = {Miller, Courtney and Choudhuri, Rudrajit and Ulloa, Mara and Haniyur, Sankeerti and DeLine, Robert and Storey, Margaret-Anne and Murphy-Hill, Emerson and Bird, Christian and Butler, Jenna L.},
title = {"Maybe We Need Some More Examples:" Individual and Team Drivers of Developer GenAI Tool Use},
year = {2026},
isbn = {9798400720253},
publisher = {Association for Computing Machinery},
address = {New York, NY, USA},
doi = {10.1145/3744916.3773155},
booktitle = {Proceedings of the 2026 IEEE/ACM 48th International Conference on Software Engineering},
pages = {2582–2595},
numpages = {14},
location = {Rio de Janeiro, Brazil},
series = {ICSE '26}
}

@inproceedings{tang2026ctf,
  author = {Tingxuan Tang and Nicolas Janis and Kalyn Asher Montague and Kevin Eykholt and Dhilung Kirat and Youngja Park and Jiyong Jang and Adwait Nadkarni and Yue Xiao},
  title = {From Assistance to Autonomy: An Empirical Study of {AI} Use in a Live {Capture-the-Flag} ({{{{{CTF}}}}}) Competition},
  booktitle = {35th USENIX Security Symposium (USENIX Security 26)},
  year = {2026},
  address = {Baltimore, MD},
  publisher = {USENIX Association},
  month = aug
}

@INPROCEEDINGS{incalmo,
  author={Singer, Brian and Lucas, Keane and Adiga, Lakshmi and Jain, Meghna and Bauer, Lujo and Sekar, Vyas},
  booktitle={2026 IEEE Symposium on Security and Privacy (SP)}, 
  title={Incalmo: an Autonomous Llm-Assisted System for Red Teaming Multi-Host Networks}, 
  year={2026},
  pages={4282-4300},
  doi={10.1109/SP63933.2026.00132}}

@inproceedings {pentestgpt,
  author = {Gelei Deng and Yi Liu and V{\'\i}ctor Mayoral-Vilches and Peng Liu and Yuekang Li and Yuan Xu and Tianwei Zhang and Yang Liu and Martin Pinzger and Stefan Rass},
  title = {{PentestGPT}: Evaluating and Harnessing Large Language Models for Automated Penetration Testing},
  booktitle = {33rd USENIX Security Symposium (USENIX Security 24)},
  year = {2024},
  isbn = {978-1-939133-44-1},
  address = {Philadelphia, PA},
  pages = {847--864},
  publisher = {USENIX Association},
  month = aug
}

@inproceedings{ullah2026cvegenie,
  author    = {Saad Ullah and
               Praneeth Balasubramanian and
               Wenbo Guo and
               Amanda Burnett and
               Hammond Pearce and
               Christopher Kruegel and
               Giovanni Vigna and
               Gianluca Stringhini},
  title     = {{CVE-Genie}: An {LLM}-Based Multi-Agent Framework for Reproducing {CVEs}},
  booktitle = {Proceedings of the 2026 ACM SIGSAC Conference on Computer and Communications Security},
  year      = {2026},
  publisher = {ACM},
}

@inproceedings{Ji2026FirmAgent,
    author    = {Ji, Jiangan and Zhang, Chao and Gan, Shuitao and Jian, Lin and Liu, Hangtian and Liu, Tieming and Zheng, Lei and Jia, Zhipeng},
    title     = {FirmAgent: Leveraging Fuzzing to Assist LLM Agents with IoT Firmware Vulnerability Discovery},
    booktitle = {Network and Distributed System Security (NDSS) Symposium},
    year      = {2026},
    month     = {February},
    pages     = {1--16},
    address   = {San Diego, CA, USA},
    doi       = {10.14722/ndss.2026.231943},
    isbn      = {979-8-9919276-8-0},
}

@inproceedings{liu2026bacagent,
  author    = {Fengyu Liu and
               Yuan Zhang and
               Zheng Lou and
               Tian Chen and
               Youkun Shi and
               Jiarun Dai and
               Enhao Li and
               Guangyu Zhou and
               Zhongfu Su and
               Zequn Fang},
  title     = {{BACAgent}: {LLM}-Powered Detection of Broken-Access-Control Vulnerabilities in Web Applications},
  booktitle = {Proceedings of the 33rd ACM Conference on Computer and Communications Security (CCS)},
  year      = {2026},
  address   = {The Hague, The Netherlands},
  publisher = {ACM}
}

@article{happe2026can,
  title={Can llms hack enterprise networks? autonomous assumed breach penetration-testing active directory networks},
  author={Happe, Andreas and Cito, J{\"u}rgen},
  journal={ACM Transactions on Software Engineering and Methodology},
  volume={35},
  number={6},
  pages={1--54},
  year={2026},
  publisher={ACM New York, NY}
}

@inproceedings{
  lin2026comparing,
  author = {Lin, Justin and Jones, Eliot and Jasper, Donovan and Ho, Ethan and Wu, Anna and Yang, Arnold and Perry, Neil and Zou, Andy and Fredrikson, Matt  and Kolter, Zico and Liang, Percy and Boneh, Dan  and Ho, Daniel},
  booktitle = {International Conference on Learning Representations},
  editor = {C. Vondrick and B. Hariharan and C. Raffel and L. Pinto and D. Yang and A. Faust},
  pages = {1872--1902},
  title = {Comparing AI Agents to Cybersecurity Professionals in Real-World Penetration Testing},
  url = {https://proceedings.iclr.cc/paper_files/paper/2026/file/0410c2ff9f872efe5a7c61a4323a5da3-Paper-Conference.pdf},
  volume = {2026},
  year = {2026}
}

@InProceedings{enigma,
  title = 	 {{E}n{IGMA}: Interactive Tools Substantially Assist {LM} Agents in Finding Security Vulnerabilities},
  author =       {Abramovich, Talor and Udeshi, Meet and Shao, Minghao and Lieret, Kilian and Xi, Haoran and Milner, Kimberly and Jancheska, Sofija and Yang, John and Jimenez, Carlos E and Khorrami, Farshad and Krishnamurthy, Prashanth and Dolan-Gavitt, Brendan and Shafique, Muhammad and Narasimhan, Karthik R and Karri, Ramesh and Press, Ofir},
  booktitle = 	 {Proceedings of the 42nd International Conference on Machine Learning},
  pages = 	 {246--355},
  year = 	 {2025},
  editor = 	 {Singh, Aarti and Fazel, Maryam and Hsu, Daniel and Lacoste-Julien, Simon and Berkenkamp, Felix and Maharaj, Tegan and Wagstaff, Kiri and Zhu, Jerry},
  volume = 	 {267},
  series = 	 {Proceedings of Machine Learning Research},
  month = 	 {13--19 Jul},
  publisher =    {PMLR},
}

@inproceedings {HLPFuzz,
author = {Yupeng Yang and Shenglong Yao and Jizhou Chen and Wenke Lee},
title = {Hybrid Language Processor Fuzzing via {LLM-Based} Constraint Solving},
booktitle = {34th USENIX Security Symposium (USENIX Security 25)},
year = {2025},
isbn = {978-1-939133-52-6},
address = {Seattle, WA},
pages = {6299--6318},
publisher = {USENIX Association},
month = aug
}

@inproceedings {PANGOLIN,
author = {Zhipeng Jia and Xiaokang Yin and Shuitao Gan and Chao Zhang and Hangtian Liu and Jiangan Ji and Enzhou Song and Ruijie Cai and Jinglei Tan and Shengli Liu},
title = {{PANGOLIN}: Fuzzing Multilingual {IoT} Firmware with {LLM-Driven} Code Analysis},
booktitle = {35th USENIX Security Symposium (USENIX Security 26)},
year = {2026},
isbn = {978-1-939133-58-8},
address = {Baltimore, MD},
pages = {673--692},
url = {https://www.usenix.org/conference/usenixsecurity26/presentation/jia-zhipeng},
publisher = {USENIX Association},
month = aug
}

@INPROCEEDINGS{VeriExploit,
  author={Wei, Chenfeng and Cai, Shiyu and Charalambous, Yiannis and Wu, Tong and Godboley, Sangharatna and Cordeiro, Lucas},
  booktitle={2025 40th IEEE/ACM International Conference on Automated Software Engineering (ASE)}, 
  title={VeriExploit: Automatic Bug Reproduction in Smart Contracts via LLMs and Formal Methods}, 
  year={2025},
  volume={},
  number={},
  pages={1415-1426},
  doi={10.1109/ASE63991.2025.00120}}

@inproceedings{hptsa,
    title = "Teams of {LLM} Agents can Exploit Zero-Day Vulnerabilities",
    author = "Zhu, Yuxuan  and
      Kellermann, Antony  and
      Gupta, Akul  and
      Li, Philip  and
      Fang, Richard  and
      Bindu, Rohan  and
      Kang, Daniel",
    editor = "Demberg, Vera  and
      Inui, Kentaro  and
      Marquez, Llu{\'i}s",
    booktitle = "Proceedings of the 19th Conference of the {E}uropean Chapter of the {A}ssociation for {C}omputational {L}inguistics (Volume 1: Long Papers)",
    month = mar,
    year = "2026",
    address = "Rabat, Morocco",
    publisher = "Association for Computational Linguistics",
    doi = "10.18653/v1/2026.eacl-long.2",
    pages = "23--35",
    ISBN = "979-8-89176-380-7",
}

@inproceedings{zeng2026pbfuzz,
  author    = {Haochen Zeng and
               Andrew Bao and
               Jiajun Cheng and
               Chengyu Song},
  title     = {{PBFuzz}: Agentic Directed Fuzzing for {PoV} Generation},
  booktitle = {Proceedings of the 33rd ACM Conference on Computer and Communications Security},
  year      = {2026},
  publisher = {ACM}
}

@misc{yang2026autoeg,
      title={AutoEG: Exploiting Known Third-Party Vulnerabilities in Black-Box Web Applications}, 
      author={Ruozhao Yang and Mingfei Cheng and Gelei Deng and Junjie Wang and Tianwei Zhang and Xiaofei Xie},
      year={2026},
      eprint={2604.00704},
      archivePrefix={arXiv},
      primaryClass={cs.CR},
}

@article{v2e,
  author = {Zhang, Jingwen and Nan, Yuhong and Ning, Kaiwen and Ye, Mingxi and Li, Wei and Xiao, Yuming and Feng, Yuming and Zhang, Weizhe and Zheng, Zibin},
  title = {V2E: Validating Smart Contract Vulnerabilities through Profit-Driven Exploit Generation and Execution},
  year = {2026},
  issue_date = {July 2026},
  publisher = {Association for Computing Machinery},
  address = {New York, NY, USA},
  volume = {3},
  number = {FSE},
  doi = {10.1145/3808108},
  journal = {Proc. ACM Softw. Eng.},
  month = jun,
  articleno = {FSE101},
  numpages = {23}
}

@inproceedings{widyasari2026vultrial,
  author    = {Ratnadira Widyasari and
               Martin Weyssow and
               Ivana Clairine Irsan and
               Han Wei Ang and
               Frank Liauw and
               Eng Lieh Ouh and
               Lwin Khin Shar and
               Hong Jin Kang and
               David Lo},
  title     = {Let the Trial Begin: A Mock-Court Approach to Vulnerability Detection using {LLM}-Based Agents},
  booktitle = {Proceedings of the 48th IEEE/ACM International Conference on Software Engineering},
  year      = {2026},
  publisher = {ACM}
}

@inproceedings {g2fuzz,
  author = {Kunpeng Zhang and Zongjie Li and Daoyuan Wu and Shuai Wang and Xin Xia},
  title = {{Low-Cost} and Comprehensive Non-textual Input Fuzzing with {LLM-Synthesized} Input Generators},
  booktitle = {34th USENIX Security Symposium (USENIX Security 25)},
  year = {2025},
  isbn = {978-1-939133-52-6},
  address = {Seattle, WA},
  pages = {6999--7018},
  publisher = {USENIX Association},
  month = aug
}

@inproceedings{shao2024nyu,
  author = {Shao, Minghao and Jancheska, Sofija and Udeshi, Meet and Dolan-Gavitt, Brendan and Xi, Haoran and Milner, Kimberly and Chen, Boyuan and Yin, Max and Garg, Siddharth and Krishnamurthy, Prashanth and Khorrami, Farshad and Karri, Ramesh and Shafique, Muhammad},
  booktitle = {Advances in Neural Information Processing Systems},
  doi = {10.52202/079017-1832},
  editor = {A. Globerson and L. Mackey and D. Belgrave and A. Fan and U. Paquet and J. Tomczak and C. Zhang},
  pages = {57472--57498},
  publisher = {Curran Associates, Inc.},
  title = {NYU CTF Bench: A Scalable Open-Source Benchmark Dataset for Evaluating LLMs in Offensive Security},
  volume = {37},
  year = {2024}
}

@inproceedings{wang2026cybergym,
 author = {WANG, ZHUN and Shi, Tianneng and He, Jingxuan and Cai, Matthew and Zhang, Jialin and Song, Dawn},
 booktitle = {International Conference on Learning Representations},
 editor = {C. Vondrick and B. Hariharan and C. Raffel and L. Pinto and D. Yang and A. Faust},
 pages = {123341--123386},
 title = {CyberGym: Evaluating {AI} Agents' Real-World Cybersecurity Capabilities at Scale},
 url = {https://proceedings.iclr.cc/paper_files/paper/2026/file/c876a2935a90aff21874e14c07dc3e33-Paper-Conference.pdf},
 volume = {2026},
 year = {2026}
}

@inproceedings{liu2026pacebench,
 author = {Liu, Zicheng and Huang, Lige and Zhang, Jie and Liu, Dongrui and Tian, Yuan and Shao, Jing},
 booktitle = {International Conference on Learning Representations},
 editor = {C. Vondrick and B. Hariharan and C. Raffel and L. Pinto and D. Yang and A. Faust},
 pages = {17137--17163},
 title = {PACEbench: A Framework for Evaluating Practical AI Cyber-Exploitation Capabilities},
 url = {https://proceedings.iclr.cc/paper_files/paper/2026/file/1cc901ebe4c549ab9222af0c1dd449b1-Paper-Conference.pdf},
 volume = {2026},
 year = {2026}
}

@InProceedings{zhu2025cvebench,
  title = 	 {{CVE}-Bench: A Benchmark for {AI} Agents’ Ability to Exploit Real-World Web Application Vulnerabilities},
  author =       {Zhu, Yuxuan and Kellermann, Antony and Bowman, Dylan and Li, Philip and Gupta, Akul and Danda, Adarsh and Fang, Richard and Jensen, Conner and Ihli, Eric and Benn, Jason and Geronimo, Jet and Dhir, Avi and Rao, Sudhit and Yu, Kaicheng and Stone, Twm and Kang, Daniel},
  booktitle = 	 {Proceedings of the 42nd International Conference on Machine Learning},
  pages = 	 {79850--79867},
  year = 	 {2025},
  editor = 	 {Singh, Aarti and Fazel, Maryam and Hsu, Daniel and Lacoste-Julien, Simon and Berkenkamp, Felix and Maharaj, Tegan and Wagstaff, Kiri and Zhu, Jerry},
  volume = 	 {267},
  series = 	 {Proceedings of Machine Learning Research},
  month = 	 {13--19 Jul},
  publisher =    {PMLR},
}

@inproceedings{zhang2025cybench,
    author = {Zhang, Andy K and Perry, Neil and Dulepet, Riya and Ji, Joey and Menders, Celeste and Lin, Justin and Jones, Eliot and Hussein, Gashon and Liu, Samantha and Jasper, Donovan and Peetathawatchai, Pura and Glenn, Ari and Sivashankar, Vikram and Zamoshchin, Daniel and Glikbarg, Leo and Askaryar, Derek and Yang, Haoxiang and Zhang, Aolin and Alluri, Rishi and Tran, Nathan and Sangpisit, Rinnara and Oseleononmen, Kenny and Boneh, Dan  and Ho, Daniel and Liang, Percy},
    booktitle = {International Conference on Learning Representations},
    editor = {Y. Yue and A. Garg and N. Peng and F. Sha and R. Yu},
    pages = {25094--25243},
    title = {Cybench: A Framework for Evaluating Cybersecurity Capabilities and Risks of Language Models},
    url = {https://proceedings.iclr.cc/paper_files/paper/2025/file/3e9412a9c1d93810ef3ef7825115016b-Paper-Conference.pdf},
    volume = {2025},
    year = {2025}
}

@inproceedings {deng2023nautilus,
author = {Gelei Deng and Zhiyi Zhang and Yuekang Li and Yi Liu and Tianwei Zhang and Yang Liu and Guo Yu and Dongjin Wang},
title = {{NAUTILUS}: Automated {RESTful} {API} Vulnerability Detection},
booktitle = {32nd USENIX Security Symposium (USENIX Security 23)},
year = {2023},
isbn = {978-1-939133-37-3},
address = {Anaheim, CA},
pages = {5593--5609},
publisher = {USENIX Association},
month = aug
}

@inproceedings {ferguson2021examining,
author = {Kimberly J. Ferguson-Walter and Maxine M. Major and Chelsea K. Johnson and Daniel H. Muhleman},
title = {Examining the Efficacy of Decoy-based and Psychological Cyber Deception},
booktitle = {30th USENIX Security Symposium (USENIX Security 21)},
year = {2021},
isbn = {978-1-939133-24-3},
pages = {1127--1144},
publisher = {USENIX Association},
month = aug
}

@misc{ptes,
  author       = {{Penetration Testing Execution Standard}},
  title        = {The Penetration Testing Execution Standard},
  howpublished = {\url{http://www.pentest-standard.org/}},
  note         = {Accessed: 2026-10-02}
}

@misc{cyber_kill_chain,
  author       = {{Lockheed Martin}},
  title        = {Cyber Kill Chain},
  howpublished = {\url{https://www.lockheedmartin.com/en-us/capabilities/cyber/cyber-kill-chain.html}},
  note         = {Accessed: 2026-09-30}
}

@InProceedings{zhuo2026ctfdojo,
  title = 	 {Training Language Model Agents to Find Vulnerabilities with {CTF}-Dojo},
  author =       {Zhuo, Terry Yue and Wang, Dingmin and Ding, Hantian and Kumar, Varun and Wang, Zijian},
  booktitle = 	 {Proceedings of the 43rd International Conference on Machine Learning},
  pages = 	 {168265--168284},
  year = 	 {2026},
  editor = 	 {Zhang, Tong and Dudik, Miroslav and Jaggi, Martin and Agarwal, Alekh and Li, Sharon and Schuurmans, Dale and Zhu, Jerry and Berkenkamp, Felix and Dong, Hanze and Bietti, Alberto},
  volume = 	 {306},
  series = 	 {Proceedings of Machine Learning Research},
  month = 	 {06--11 Jul},
  publisher =    {PMLR}
}

@inproceedings{intercode,
 author = {Yang, John and Prabhakar, Akshara and Narasimhan, Karthik and Yao, Shunyu},
 booktitle = {Advances in Neural Information Processing Systems},
 doi = {10.52202/075280-1035},
 editor = {A. Oh and T. Naumann and A. Globerson and K. Saenko and M. Hardt and S. Levine},
 pages = {23826--23854},
 publisher = {Curran Associates, Inc.},
 title = {InterCode: Standardizing and Benchmarking Interactive Coding with Execution Feedback},
 volume = {36},
 year = {2023}
}

@inproceedings{shao2026ctftiny,
  author    = {Minghao Shao and Nanda Rani and Kimberly Milner and
               Haoran Xi and Meet Udeshi and Saksham Aggarwal and
               Venkata Sai Charan Putrevu and Sandeep K. Shukla and
               Prashanth Krishnamurthy and Farshad Khorrami and
               Ramesh Karri and Muhammad Shafique},
  title     = {Towards Effective Offensive Security LLM Agents:
               Hyperparameter Tuning, LLM as a Judge, and a
               Lightweight CTF Benchmark},
  booktitle = {Proceedings of the AAAI Conference on Artificial Intelligence},
  volume    = {40},
  number    = {35},
  pages     = {29660--29668},
  year      = {2026},
  doi       = {10.1609/aaai.v40i35.40210}
}

@inproceedings{autopenbench,
    title = "{A}uto{P}en{B}ench: A Vulnerability Testing Benchmark for Generative Agents",
    author = "Gioacchini, Luca  and
      Delsanto, Alexander  and
      Drago, Idilio  and
      Mellia, Marco  and
      Siracusano, Giuseppe  and
      Bifulco, Roberto",
    editor = "Potdar, Saloni  and
      Rojas-Barahona, Lina  and
      Montella, Sebastien",
    booktitle = "Proceedings of the 2025 Conference on Empirical Methods in Natural Language Processing: Industry Track",
    month = nov,
    year = "2025",
    address = "Suzhou (China)",
    publisher = "Association for Computational Linguistics",
    doi = "10.18653/v1/2025.emnlp-industry.114",
    pages = "1615--1624",
    ISBN = "979-8-89176-333-3"
}

@article{ctfedu,
title = {Cybersecurity knowledge and skills taught in capture the flag challenges},
journal = {Computers \& Security},
volume = {102},
pages = {102154},
year = {2021},
issn = {0167-4048},
doi = {https://doi.org/10.1016/j.cose.2020.102154},
author = {Valdemar Švábenský and Pavel Čeleda and Jan Vykopal and Silvia Brišáková},
}

@INPROCEEDINGS{singh2026soc,
  author={Singh, Ronal and Tariq, Shahroz and Jalalvand, Fatemeh and Chhetri, Mohan Baruwal and Nepal, Surya and Paris, Cecile and Lochner, Martin},
  booktitle={2026 IEEE Symposium on Security and Privacy (SP)}, 
  title={LLMs in the SOC: An Empirical Study of Human-AI Collaboration in Security Operations Centres}, 
  year={2026},
  pages={4262-4281},
  doi={10.1109/SP63933.2026.00111}
}

@article{happe2026llms,
author = {Happe, Andreas and Kaplan, Aaron and Cito, J{\"u}rgen},
title = {LLMs as Hackers: Autonomous Linux Privilege Escalation Attacks},
year = {2026},
issue_date = {Feb 2026},
publisher = {Kluwer Academic Publishers},
address = {USA},
volume = {31},
number = {3},
issn = {1382-3256},
doi = {10.1007/s10664-025-10758-3},
journal = {Empirical Softw. Engg.},
month = feb,
numpages = {50}
}

@inproceedings{roch2024navigating,
  title={Navigating autonomy: Unveiling security experts' perspectives on augmented intelligence in cybersecurity},
  author={Roch, Neele and Sievers, Hannah and Sch{\"o}ni, Lorin and Zimmermann, Verena},
  booktitle={Twentieth Symposium on Usable Privacy and Security (SOUPS 2024)},
  pages={41--60},
  year={2024}
}

@article{bauer2026whitebook,
author = "Lujo Bauer and David Brumley and Joseph Calandrino and Nicolas Christin and Giulia Fanti and Virgil Gligor and Bryan Parno and Jignesh Patel and Vyas Sekar and Justine Sherry Martins",
title = "{Creating a Scientific Foundation for Cyber Autonomy}",
year = "2026",
month = "3",
url = "https://kilthub.cmu.edu/articles/report/Creating_a_Scientific_Foundation_for_Cyber_Autonomy/31769038",
doi = "10.1184/R1/31769038.v1"
}

@misc{rapid7severity,
  author       = {{Rapid7}},
  title        = {Review Vulnerability Details},
  howpublished = {Rapid7 Application Security Documentation},
  url          = {https://docs.rapid7.com/insightappsec/review-vulnerabilities/},
  note         = {Accessed: 2026-10-15}
}

@inproceedings {akgul2023bugbounty,
author = {Omer Akgul and Taha Eghtesad and Amit Elazari and Omprakash Gnawali and Jens Grossklags and Michelle L. Mazurek and Daniel Votipka and Aron Laszka},
title = {Bug {Hunters{\textquoteright}} Perspectives on the Challenges and Benefits of the Bug Bounty Ecosystem},
booktitle = {32nd USENIX Security Symposium (USENIX Security 23)},
year = {2023},
isbn = {978-1-939133-37-3},
address = {Anaheim, CA},
pages = {2275--2291},
url = {https://www.usenix.org/conference/usenixsecurity23/presentation/akgul},
publisher = {USENIX Association},
month = aug
}

@online{stenberg2025death,
  author = {Daniel Stenberg},
  title  = {Death by a Thousand Slops},
  year   = {2025},
  month  = jul,
  day    = {14},
  url    = {https://daniel.haxx.se/blog/2025/07/14/death-by-a-thousand-slops/},
  note   = {July 14, 2025}
}

@inproceedings{syros2026saga,
  author = {Georgios Syros and Anshuman Suri and Jacob Ginesin and Cristina Nita-Rotaru and Alina Oprea},
  title = {{SAGA}: A Security Architecture for Governing {AI} Agentic Systems},
  booktitle = {Network and Distributed System Security Symposium (NDSS)},
  year = {2026},
  doi = {10.14722/ndss.2026.230869},
  url = {https://www.ndss-symposium.org/ndss-paper/saga-a-security-architecture-for-governing-ai-agentic-systems/}
}

@inproceedings{klees2018evaluating,
  author = {Klees, George and Ruef, Andrew and Cooper, Benji and Wei, Shiyi and Hicks, Michael},
  title = {Evaluating Fuzz Testing},
  year = {2018},
  isbn = {9781450356930},
  publisher = {Association for Computing Machinery},
  address = {New York, NY, USA},
  url = {https://doi.org/10.1145/3243734.3243804},
  doi = {10.1145/3243734.3243804},
  booktitle = {Proceedings of the 2018 ACM SIGSAC Conference on Computer and Communications Security},
  pages = {2123–2138},
  numpages = {16},
  location = {Toronto, Canada},
  series = {CCS '18}
}

@misc{kim2026ledger,
      title={LEDGER: Claim-to-Evidence Trace Graphs for Auditing LLM Agents}, 
      author={Daehong Kim and Haichao Miao and Shusen Liu},
      year={2026},
      eprint={2608.18398},
      archivePrefix={arXiv},
      primaryClass={cs.HC},
      url={https://arxiv.org/abs/2608.18398}, 
}

@inproceedings{stechly2025selfverification,
 author = {Stechly, Kaya and Valmeekam, Karthik and Kambhampati, Subbarao},
 booktitle = {International Conference on Learning Representations},
 editor = {Y. Yue and A. Garg and N. Peng and F. Sha and R. Yu},
 pages = {98190--98243},
 title = {On the self-verification limitations of large language models on reasoning and planning tasks},
 url = {https://proceedings.iclr.cc/paper_files/paper/2025/file/f3c5e56274140e0420baa3916c529210-Paper-Conference.pdf},
 volume = {2025},
 year = {2025}
}

@article{satter2026aibugs,
  author  = {Raphael Satter and A. J. Vicens},
  title   = {AI-Discovered Bugs May Soon Trigger Fresh Disruptions, Cyber Experts Say},
  journal = {Reuters},
  year    = {2026},
  month   = oct,
  day     = {23},
  url     = {https://www.reuters.com/legal/litigation/ai-discovered-bugs-may-soon-trigger-fresh-disruptions-cyber-experts-say-2026-10-23/},
  note    = {Accessed: 2026-10-26}
}

@inproceedings {guilhem2025attacker,
author = {Guilhem Lacombe and S{\'e}bastien Bardin},
title = {Attacker Control and Bug Prioritization},
booktitle = {34th USENIX Security Symposium (USENIX Security 25)},
year = {2025},
isbn = {978-1-939133-52-6},
address = {Seattle, WA},
pages = {4541--4560},
url = {https://www.usenix.org/conference/usenixsecurity25/presentation/lacombe},
publisher = {USENIX Association},
month = aug
}

@inproceedings {li2026patchweaver,
author = {Rui Li and Shuang Cao},
title = {{PatchWeaver}: {Risk-Bounded} Autonomous Vulnerability Remediation Under {Change-Management} Policies},
booktitle = {35th USENIX Security Symposium (USENIX Security 26)},
year = {2026},
isbn = {978-1-939133-58-8},
address = {Baltimore, MD},
pages = {6839--6858},
url = {https://www.usenix.org/conference/usenixsecurity26/presentation/li-rui},
publisher = {USENIX Association},
month = aug
}

@INPROCEEDINGS{ayala2025bugbounty,
  author={Ayala, Jessy and Ngo, Steven and Garcia, Joshua},
  booktitle={2025 IEEE Symposium on Security and Privacy (SP)}, 
  title={A Deep Dive into How Open-Source Project Maintainers Review and Resolve Bug Bounty Reports}, 
  year={2025},
  volume={},
  number={},
  pages={522-538},
  doi={10.1109/SP61157.2025.00063}}

\appendices
\section{Survey on LLM-Assisted Penetration Testing}
\label{app:survey}

We distinguish the full survey sample ($n=167$) from respondents reporting LLM use in PT during the preceding 12 months ($n=159$). For activity-specific items, percentages are calculated among applicable responses. Summaries spanning all six PT activities include respondents with applicable answers for every activity. Collaboration-mode comparisons group respondents by their self-reported primary mode. Below, we provide the core definitions used in Section~\ref{sec:rq1}.

\subsection{PT Activities}

\noindent\textbf{Information gathering and attack-surface analysis.}
Discovering and organizing assets, hosts, ports, services, technology stacks, and potential attack entry points.

\noindent\textbf{Vulnerability analysis and attack-direction selection.}
Analyzing potential vulnerabilities, forming attack hypotheses, and determining subsequent testing directions.

\noindent\textbf{Exploit implementation.}
Generating, writing, modifying, adapting, or executing payloads, proof-of-concept code, exploits, attack commands, and scripts, and resolving technical implementation problems.

\noindent\textbf{Post-exploitation and attack-chain extension.}
Using established access, identities, credentials, privileges, or other acquired resources to obtain further control or extend an attack chain.

\noindent\textbf{Vulnerability and impact verification.}
Confirming vulnerabilities and exploitation results, and assessing privileges, affected scope, and security consequences.

\noindent\textbf{Result organization and reporting.}
Organizing evidence, vulnerability descriptions, reproduction steps, impact assessments, remediation recommendations, and PT reports.

\subsection{Human--LLM Collaboration Modes}

\noindent\textbf{Local assistance.}
The human analyzes the target, determines subsequent steps, and conducts testing; the LLM provides localized assistance such as information retrieval, interpretation, commands, payloads, or code.

\noindent\textbf{Subtask delegation.}
The human determines a specific task and delegates it to the LLM; after completion, the human determines subsequent testing directions.

\noindent\textbf{Joint decision making.}
The human and LLM jointly analyze the environment and develop or revise attack hypotheses and plans through repeated interaction.

\noindent\textbf{Goal-level delegation.}
The human specifies testing goals, scope, and constraints; the LLM determines subsequent directions, plans multiple steps, and executes them continuously, while the human supervises and intervenes as needed.

\subsection{Practical Constraints}

\noindent\textbf{Capability and reliability limitations.}
Incorrect judgments, ineffective suggestions, or unstable results that make LLM outputs difficult to trust.

\noindent\textbf{Task suitability limitations.}
PT tasks for which using an LLM provides little advantage over direct human execution.

\noindent\textbf{Data and compliance limitations.}
Restrictions on sharing target information, vulnerability data, or internal materials because of organizational or confidentiality requirements.

\noindent\textbf{Execution-environment and tool-access limitations.}
Difficulty directly accessing or continuously operating target environments, jump hosts, proxies, credentials, sessions, or existing security tools.

\noindent\textbf{Resource and service limitations.}
Constraints involving API or token costs, quotas, context length, response latency, compute resources, or long-horizon tasks.

\noindent\textbf{No noticeable limitations.}
No major factor currently limits LLM use in penetration testing.

\subsection{Reported Problem Categories}

\noindent\textbf{Incorrect vulnerability or technical judgments.}
Incorrect judgments about vulnerabilities, software versions, CVEs, or other technical facts.

\noindent\textbf{Unreliable attack implementation.}
Generated payloads, PoCs, or exploits fail to function correctly, reproduce actual exploitation results, or match an exploitable path.

\noindent\textbf{Environment or attack-direction misunderstanding.}
Misinterpreting the target environment or tool output, or forming incorrect vulnerability hypotheses or subsequent directions.

\noindent\textbf{Insufficient handling of complex environments and state.}
Difficulty handling jump hosts, proxies, network paths, authentication state, credentials, sessions, or state continuity across multiple steps or systems.

\noindent\textbf{Scope or safety-boundary deviation.}
Suggesting or attempting actions beyond the testing scope, expanding attack targets or privileges, or performing potentially destructive or high-risk operations.

\noindent\textbf{Few or no noticeable problems.}
Few or no significant problems are encountered when using LLMs for PT.

\section{Participant Matching and Group Assignment}
\label{app:participants}

\textbf{Experience Characterization.}
Before group assignment, we characterize each participant's prior penetration testing experience using their resumes and documented hands-on project experience. When available, public technical profiles and project records are used as supplementary evidence. We categorize each participant's penetration testing experience as R4 (extensive real-world PT/red-team experience), R3 (sustained independent PT experience), R2 (established hands-on PT experience), and R1 (practical PT project experience). Participants' LLM use experience is considered as an auxiliary factor during matching.

\textbf{Matching and Assignment.}
PT experience serves as the primary matching criterion. Participants with similar PT backgrounds are paired, with participants in the same experience category matched whenever possible. When multiple suitable matches are available, we further consider the similarity of their PT backgrounds and LLM-use experience. Each matched pair is then randomly split across the \textit{LLM-Available} and \textit{Human-Only} conditions. As shown in Table~\ref{tab:participant-balance}, the final sample contains 20 participants in each condition.

\begin{table}[t]
\centering
\caption{Pre-study PT experience by condition.}
\label{tab:participant-balance}
\begin{tabular}{lcc}
\toprule
\textbf{PT Experience} &
\textbf{LLM-Available} &
\textbf{Human-Only} \\
\midrule
R4: Extensive & 3 & 3 \\
R3: Sustained & 10 & 10 \\
R2: Established & 6 & 6 \\
R1: Practical & 1 & 1 \\
\midrule
\textbf{Total} & \textbf{20} & \textbf{20} \\
\bottomrule
\end{tabular}
\end{table}

\begin{figure}[t]
    \centering
    \includegraphics[width=\columnwidth]{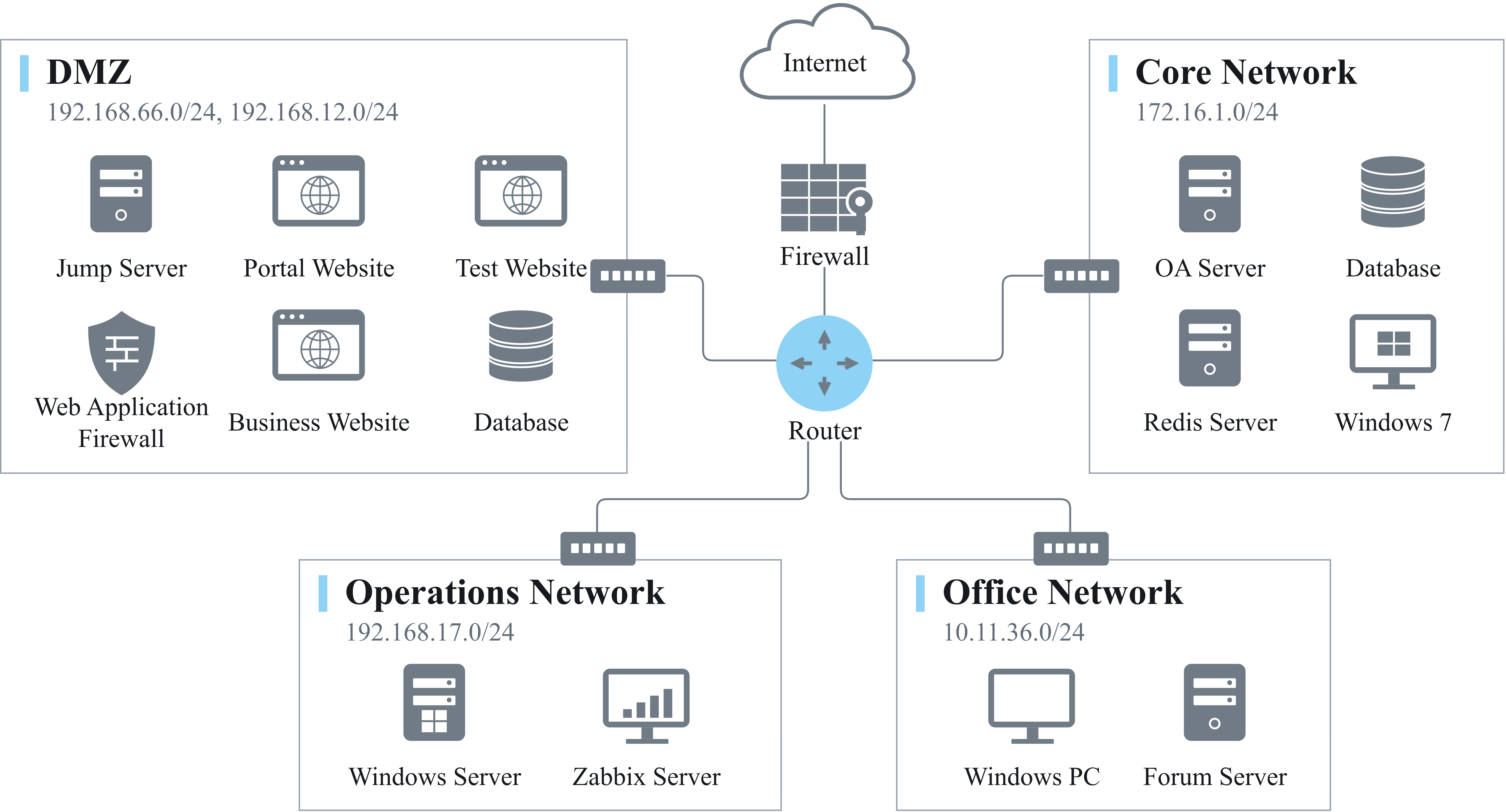}
    \caption{Network topology of the penetration testing environment.}
    \label{fig:topology}
\end{figure}

\section{Penetration Testing Environment}
\label{app:environment}

Figure~\ref{fig:topology} shows the penetration testing environment used in this study. The environment consists of four network zones: a DMZ, an office network, an operations network, and a core network. Only the DMZ exposes public-facing entry points, while the office, operations, and core networks are internal segments isolated by the firewall that become reachable only through further penetration and pivoting. The environment hosts heterogeneous targets, including Web applications, databases, Windows and Linux hosts, operational management services, and internal business applications. A Web application firewall (WAF) is deployed in the DMZ.

Participants perform the assessment in a black-box setting. As target information, they receive only the initial entry IP address and authorization scope; the internal topology, hosts, services, and vulnerabilities are not disclosed. Participants therefore need to discover reachable assets and services during testing and decide which attack paths to pursue as new information and access become available.

\section{Human--LLM Interaction Coding Scheme}
\label{app:interaction-coding}

We code each substantive human intervention. PT activity identifies the primary activity underway at the intervention, following the six categories defined in Appendix~\ref{app:survey}.

We distinguish six intervention types according to the intents expressed in the human input. Multiple labels are assigned when an intervention expresses distinct intents:

\noindent\textit{\textbf{APPROVE.}} Accepting or authorizing an LLM-proposed action or plan, or permitting existing work to continue.

\noindent\textit{\textbf{STEER.}} Specifying or revising a target, method, attack direction, priority, or next step.

\noindent\textit{\textbf{CHECK.}} Requesting information, a status update, an explanation, or verification.

\noindent\textit{\textbf{CORRECT.}} Identifying an error or inconsistency in the LLM's reasoning, actions, or statements.

\noindent\textit{\textbf{INFORM.}} Providing task-relevant information, including observations, tool results, environmental context, or reporting information.

\noindent\textit{\textbf{CONSTRAIN.}} Setting limits on targets or actions, or interrupting or terminating execution.

Intervention type is determined from the human input and its preceding context. We distinguish three attack-path responses based on the LLM's subsequent response and observed actions, up to the next human intervention or the end of the session:

\noindent\textit{\textbf{CONTINUE.}} Maintaining or resuming the existing attack path, including temporary suspension with the path preserved.

\noindent\textit{\textbf{EXPAND.}} Adding a substantive testing direction while retaining the existing attack path.

\noindent\textit{\textbf{REDIRECT.}} Leaving the current attack path, and switching to another direction.

We additionally classify interventions as proactive when humans take the initiative to provide input, or reactive when the LLM pauses and waits for a human response.

\section{Participant-Level Score Composition}
\label{app:scores}

\begin{figure*}[t]
    \centering
    \includegraphics[width=\textwidth]{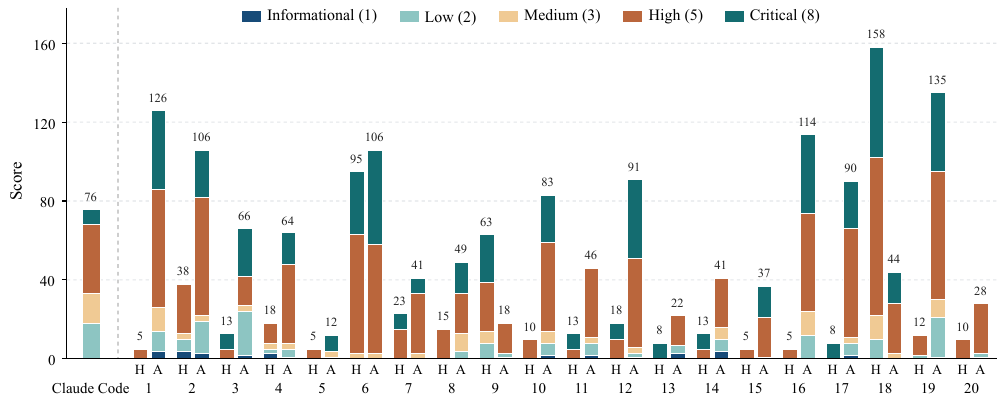}
    \caption{Score composition by finding severity for the autonomous agent and each participant (H: Human-Only; A: LLM-Available).}
    \label{fig:rq2-score-composition}
\end{figure*}

Figure~\ref{fig:rq2-score-composition} presents individual score compositions underlying the comparisons in Section~\ref{sec:rq2-overall}. The leftmost bar represents the autonomous agent, followed by the 20 matched participant pairs. Within each pair, the \textit{Human-Only} participant (H) appears before the \textit{LLM-Available} participant (A). Each colored segment shows the points contributed by findings at one severity level, following the scoring scheme in Table~\ref{tab:severity}. Numbers above the bars indicate total scores, and numbers below the pairs identify the matches used in the paired comparisons.

\section{LLM Usage}
\label{app:llm-usage}

\begin{figure}[t]
    \centering
    \includegraphics[width=\columnwidth]{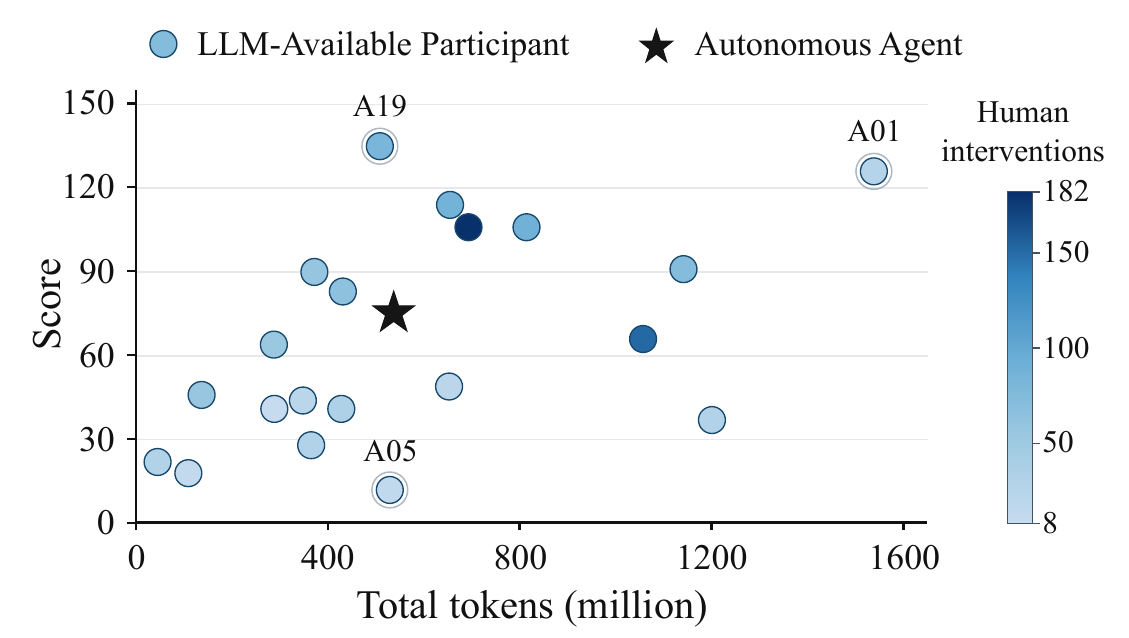}
    \caption{LLM total token consumption and scores. Color indicates human intervention count, and the star denotes the autonomous agent.}
    \label{fig:llm-usage}
\end{figure}

We measure LLM usage through total token consumption and human involvement through the 1,131 coded interventions analyzed in Section~\ref{sec:rq3}. As shown in Figure~\ref{fig:llm-usage}, token consumption ranges from 43.9 million to 1,536.8 million per participant (median: 468.6 million), while intervention counts range from 8 to 182 (median: 44). Across the 20 participants, token consumption is positively correlated with scores (Spearman's $\rho=0.54$, $p=0.015$), while intervention counts are positively correlated with both scores ($\rho=0.70$, $p<0.001$) and finding counts ($\rho=0.74$, $p<0.001$). Token consumption has a modest, statistically non-significant association with intervention counts ($\rho=0.42$, $p=0.065$).

Although all 20 participants in the \textit{LLM-Available} condition have access to the same LLM configuration, their actual usage varies substantially. Participants with similar token consumption can exhibit substantial differences in performance and intervention counts. For example, A05 and A19 each consume approximately 500 million tokens but record 14 and 83 interventions and obtain scores of 12 and 135, respectively. A01 achieves a comparable score of 126 with 27 interventions, while consuming roughly three times as many tokens as A19. These differences suggest that neither higher token consumption nor more frequent human intervention guarantees better PT outcomes. Therefore, usage intensity alone does not adequately reflect the effectiveness of LLM use in PT.

\end{document}